\documentclass[pra,longbibliography,twocolumn,showpacs,nofootinbib,superscriptaddress,notitlepage,accepted=2019-05-14]{quantumarticle}
\pdfoutput=1
\usepackage{amsmath}
\usepackage{amssymb,bm}
\usepackage{amsthm}
\usepackage[utf8]{inputenc}
\usepackage[english]{babel}
\usepackage[T1]{fontenc}
\usepackage{tikz}

\newtheorem{innercustomgeneric}{\customgenericname}
\providecommand{\customgenericname}{}
\newcommand{\newcustomtheorem}[2]{%
  \newenvironment{#1}[1]
  {%
   \renewcommand\customgenericname{#2}%
   \renewcommand\theinnercustomgeneric{##1}%
   \innercustomgeneric
  }
  {\endinnercustomgeneric}
}

\newcustomtheorem{customthm}{Theorem}
\newcustomtheorem{definitionBob}{Definition}

\usepackage{color,dsfont} 
\usepackage{graphicx}
\usepackage{ragged2e}
\usepackage[colorlinks=true, hyperindex, breaklinks, linkcolor=blue, urlcolor=blue, citecolor=blue]{hyperref} 
\usepackage[normalem]{ulem}
\usepackage[capitalise]{cleveref}
\usepackage{mathrsfs}
\usepackage[caption=false]{subfig}
\usepackage{mathtools}
\usepackage{float}
\usepackage{verbatim}
\usepackage{latexsym}
\usepackage{amsmath}
\usepackage{amssymb}
\usepackage{setspace}
\usepackage{amsfonts}
\usepackage{stmaryrd}
\usepackage{xcolor}
\usepackage{enumitem}
\usepackage{lipsum}
\DeclarePairedDelimiter{\ceil}{\lceil}{\rceil}

\newcommand{\cP}{\mathcal{P}}

 \newtheorem{definition}{Definition}

\newcommand{\ket}[1]{|#1\rangle}  
 
\newcommand{\codepar}[1]{\ensuremath{[\![#1]\!]}}

\begin{document}

\title{Fault-tolerant magic state preparation with flag qubits}

\author{Christopher Chamberland}
\email{christopher.chamberland@ibm.com}
\affiliation{
   IBM T. J. Watson Research Center,
    Yorktown Heights, NY, 10598, United States
    }
\affiliation{
   Institute for Quantum Computing and Department of Physics and Astronomy,
    University of Waterloo,
    Waterloo, Ontario, N2L 3G1, Canada
    }
    
    \author{Andrew Cross}
\email{awcross@us.ibm.com}
\affiliation{
   IBM T. J. Watson Research Center,
    Yorktown Heights, NY, 10598, United States
    }

\begin{abstract}
Magic state distillation is one of the leading candidates for implementing universal fault-tolerant logical gates. However, the distillation circuits themselves are not fault-tolerant, so there is additional cost to first implement encoded Clifford gates with negligible error. In this paper we present a scheme to fault-tolerantly and directly prepare magic states using flag qubits. One of these schemes requires only three ancilla qubits, even with noisy Clifford gates. We compare the physical qubit and gate cost of our scheme to the magic state distillation protocol of Meier, Eastin, and Knill (MEK), which is efficient and uses a small stabilizer circuit. For low enough noise rates, we show that in some regimes the overhead can be improved by several orders of magnitude compared to the MEK scheme which uses Clifford operations encoded in the codes considered in this work. 
\end{abstract}


\maketitle

\section{Introduction}
\label{sec:Intro}

Certain algorithms can be implemented efficiently on quantum computers, whereas the best known classical algorithms require superpolynomial resources \cite{Shor94, algzoo}. At present, however, quantum devices are dramatically noisier then their classical counterparts. For all but the shortest depth quantum computations to succeed with high probability, operations will need to be performed with very low failure rates. Fault-tolerant quantum error correction provides one way to achieve the desired logical failure rates.

A straightforward way to implement fault-tolerant logical gates is to use transversal operations. However by the Eastin-Knill theorem, for any error correcting code, there will always be at least one gate in a universal gate set which cannot be implemented fault-tolerantly using transversal operations \cite{EK09}. In recent years, many proposals for fault-tolerant universal gate constructions have been introduced \cite{KLZ96,JL14,PR13,ADP14,Bombin14,YTC16}. One of the earliest proposals, known as magic state distillation, uses resource states that, along with stabilizer operations, can simulate a non-Clifford operation \cite{Knill04,BK05}. These resource states are called magic states since they can also be distilled using stabilizer operations. Despite considerable effort in alternative schemes, magic state distillation remains a leading candidate for universal fault-tolerant quantum computation. 

Recently, very efficient distillation protocols have been developed which require substantially fewer resource states to achieve a desired target error rate of the magic state being distilled \cite{Haah2017magicstate,Haah2018codesprotocols,SubLogOverHaahHastings2018,InteSizeStateDist}. When studying magic state distillation protocols, it is often assumed that all Clifford operations can be implemented perfectly and that only the resource states can introduce errors into the protocol. However, with current quantum devices, two qubit gates are amongst the noisiest components of the circuits. In practice, Clifford gates with very low failure rates can be achieved by performing encoded versions of the gates in large enough error correcting codes, such that the failure rates of the Clifford operations are negligible compared to the non-Clifford components of the circuit. However the overhead cost associated with performing magic state distillation with encoded Clifford operations is often not considered (there are exceptions such as \cite{FMMC12,GotoMagicState, OC17, L18}). If the overhead cost of encoded Clifford operations is taken into account, it is not clear that magic state distillation schemes which minimize the resource state cost would achieve the lowest overhead when used in a quantum algorithm. 

Recently, new schemes using flag qubits have been introduced to implement error correction protocols using the minimal number of ancilla qubits to measure the codes stabilizer generators \cite{YK17, CR17v1,CR17v2,CB17,TCD18Flag,ReichardtFlag18}. The idea behind flag error correction is to use extra ancilla qubits which flag when $v \le t$ faults result in a data qubit error of weight greater then $v$ (here $t = \lfloor (d-1)/2 \rfloor$ where $d$ is the distance of the code). The flag information can then be used in addition to the error syndrome to deduce the data qubit error. 

In this paper, we propose a new scheme to fault-tolerantly prepare magic states which requires a minimal amount of extra ancilla qubits (in our case only one extra qubit). In particular, we consider a full circuit level noise model (which includes noisy Clifford operations) where gates can be applied between any pair of qubits. In this model, we compare the overhead cost of our scheme to a magic state distillation scheme introduced by Meier, Eastin and Knill (MEK), due to its efficiency and small circuit size \cite{MEK12}. Using the same encoded Clifford operations in both schemes, we show that in some regimes the qubit and gate overhead cost of our scheme can be smaller by several orders of magnitude as compared to the MEK scheme. We note that currently, state of-the-art surface-code-based magic state distillation schemes are given in \cite{FowlerGidney1,FowlerGidney2}. Furthermore, an efficient state injection scheme using the surface code is given in \cite{LiMagicState17}. The surface-code-based magic state distillation schemes have been shown to achieve logical error rates of $10^{-11}$ for physical error rates of $10^{-3}$ and can thus tolerate larger physical error rates than the scheme proposed in this work. On the other hand, in low noise rate regimes ($10^{-5} \le p \le 10^{-4}$), we show that the qubit and gate overhead cost of our scheme is lower compared to distance-three surface code implementations of magic state distillation. For larger distances, a more thorough analysis of surface code implementations of magic state distillation for error rates mentioned above is required to determine if our scheme has a smaller overhead. The scheme we numerically analyze in this work does require code concatenation for distances $d>3$ and thus requires non-local connectivity\footnote{In addition we show how our scheme can be extended to higher distance color codes, but these are not numerically analyzed in this work.}. However, the goal of this work is not solely to outperform the surface code, but instead to provide alternative magic state preparation schemes with low overhead. Indeed the proposed scheme in this work could be accessible to near term ion trap based architectures which could be amongst the first experiments to demonstrate the fault-tolerant implementation of logical non-Clifford gates. 

We point out that a fault-tolerant magic state preparation scheme was previously considered by Aliferis, Gottesman and Preskill (AGP) in \cite{AGP06}. However, the scheme proposed by AGP requires the preparation and verification of large cat state ancillas to perform the logical measurements. Furthermore, Steane error correction, which requires a large number of extra ancilla qubits, was used for the error correction circuits. Our scheme does not require the preparation of large ancilla states, uses smaller error correction circuits, and has an improved threshold compared to the scheme proposed by AGP.

The paper is structured as follows. In \cref{sec:NoiseModelNotation} we introduce the basic notation and noise model that we used throughout the remainder of the manuscript. In \cref{sec:SteaneCode} we describe our scheme for fault-tolerantly preparing magic states. We provide both an error detecting and and correcting scheme. Proofs of fault-tolerance are given in \cref{app:FaultToleranceProofs}. In \cref{sec:MEKdistillationCircuits} we briefly review the MEK magic state distillation scheme and in \cref{sec:Discussion} we compare the qubit and gate overhead costs of both schemes. Details of the overhead and numerical analysis are provided in \cref{app:HprepOverheadAnalysis,app:MEKOverheadAnalysis,app:StateVectorSim}.

\section{Basic notation and noise model}
\label{sec:NoiseModelNotation}

We define $\cP_n^{(1)}$ to be the $n$-qubit Pauli group (i.e. $n$-fold tensor products of the identity $I$ and the three Pauli matrices $X,\ Y,$ and $Z$ and all scalar multiples of them by $\pm 1$ and $\pm i$). The Clifford group is defined by $\cP_n^{(2)} = \{U : \forall P \in \cP_n^{(1)}, UPU^\dagger \in \cP_n^{(1)}\}$ and is generated by the single qubit Hadamard, $Y\Big( \frac{\pi}{2} \Big) = e^{-i\frac{\pi}{4}Y}$ gates
\begin{align}
H = \frac{1}{\sqrt{2}} \left( \begin{array}{cc}
                                          1 & 1 \\
                                          1 &-1\\                                                                     
                                          \end{array} \right),
\ \ {\rm and}\ \  
Y\Big( \frac{\pi}{2} \Big) = \frac{1}{\sqrt{2}} \left( \begin{array}{cc}
                                          1 & -1 \\
                                          1 &1\\                                                                     
                                          \end{array} \right),
\label{eq:HOp}
\end{align}
as well as $Z(\frac{\pi}{2}) = e^{-i\frac{\pi}{4}Z}$ and the two-qubit CNOT gate (which we write as $C_{X}$) with
\begin{align}
C_{X} \ket{a}\ket{b} = \ket{a} \ket{a \oplus b},
\label{eq:CNOTgate}                                          
\end{align}
where $\ket{a}$ and $\ket{b}$ are computational basis states. In general, a controlled-$U$ gate is
\begin{align}
C_{U} = \frac{1}{2}(I+Z) \otimes I + \frac{1}{2}(I-Z) \otimes U.
\end{align}

\begin{figure}[H]
	\centering
	\includegraphics[width=0.4\textwidth]{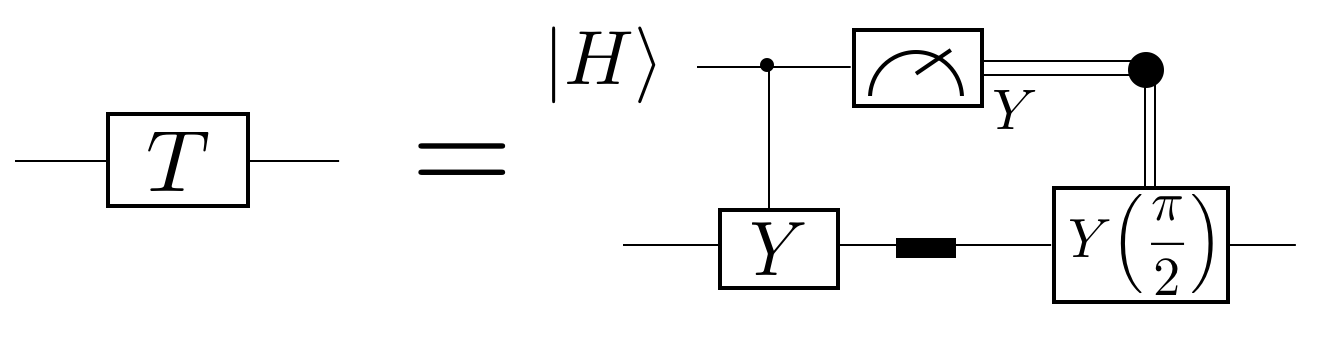}
	\caption{Circuit for simulating a $T$ gate using an $\ket{H}$ state along with $Y\big( \frac{\pi}{2} \big)$ and $C_{Y}$ gates and a measurement in the $Y$-basis. The $Y\big( \frac{\pi}{2} \big)$ gate is only applied if the measurement outcome is $+1$. The black rectangular box represents an idle location. The circuit for simulating $T^{\dagger}$ is obtained by replacing $Y\big( \frac{\pi}{2} \big)$ with $Y\big( -\frac{\pi}{2} \big)$ and applying this gate if the measurement outcome is $-1$.}
	\label{fig:TgateCirc}
\end{figure}

The state
\begin{align}
\ket{H} = \cos{\frac{\pi}{8}}\ket{0} + \sin{\frac{\pi}{8}}\ket{1} = T\ket{0},
\end{align}
is the $+1$ eigenstate of the Hadamard operator and
\begin{align}
T = e^{-i\frac{\pi Y}{8}} =  \left( \begin{array}{cc}
                                          \cos{\frac{\pi}{8}} &  -\sin{\frac{\pi}{8}} \\
                                          \sin{\frac{\pi}{8}} & \cos{\frac{\pi}{8}}\\                                                                     
                                          \end{array} \right)
\end{align}
 does not belong to the Clifford group\footnote{Note that some references define the $T$ gate as $T = \frac{e^{i\pi/4}}{\sqrt{2}}\left( \begin{array}{cc}
                                          1 & 1 \\
                                          i & -i\\                                                                     
                                          \end{array} \right)$ which is a Clifford gate, while others define $T=\mathrm{diag}(1,e^{i\pi/4})$ which is non-Clifford but Clifford-equivalent to the gate we have defined as $T$.}. We also write 
\begin{align}
\ket{-H} = Y\ket{H}
\end{align}
as the $-1$ eigenstate of the Hadamard operator.

The $\ket{H}$ state is an example of a magic state. Magic states can be distilled to produce reliable states from a larger number of noisy copies using only stabilizer operations \cite{BK05}. The reliable magic states can be used, together with stabilizer operations, as resource states for universal quantum computation. In particular, as shown in \cref{fig:TgateCirc}, the $\ket{H}$ state, along with $Y\Big( \frac{\pi}{2} \Big)$, $C_{Y}$, and a measurement in the $Y$-basis, can be used to simulate a $T$ gate. Note that some schemes choose to distill the state
\begin{align}
\ket{A_{\frac{\pi}{4}}} \equiv \frac{1}{\sqrt{2}}(\ket{0} + e^{i \frac{\pi}{4}}\ket{1}) = e^{i \frac{\pi}{8}}HS^{\dagger}\ket{H},
\label{eq:Api4State}
\end{align}
which is equivalent to the $\ket{H}$ state up to products of Clifford gates. For instance, see the  $\ket{A_{\frac{\pi}{4}}}$ distillation scheme in \cite{BK05}.

Throughout this paper, we will use the following circuit-level noise model in all simulations:
 
\begin{enumerate}
        \item With probability $p$, each single-qubit gate location is followed by a Pauli error drawn uniformly and independently from $\{ X,Y,Z \}$.
	\item With probability $p$, each two-qubit gate is followed by a two-qubit Pauli error drawn uniformly and independently from $\{I,X,Y,Z\}^{\otimes 2}\setminus \{I\otimes I\}$.	
	\item With probability $\frac{2p}{3}$, the preparation of the $\ket{0}$ state is replaced by $\ket{1}=X\ket{0}$. Similarly, with probability $\frac{2p}{3}$, the preparation of the $\ket{+}$ state is replaced by $\ket{-}=Z\ket{+}$.
	\item With probability $p$, the preparation of the $\ket{H}$ state is replaced by $P\ket{H}$ where $P$ is a Pauli error drawn uniformly and independently from $\{ X,Y,Z \}$.
	\item With probability $\frac{2p}{3}$, any single qubit measurement has its outcome flipped. 	
	\item Lastly, with probability $p/100$, each idle gate location is followed by a Pauli error drawn uniformly and independently from $\{ X,Y,Z \}$.
\end{enumerate}

Our assumption of $p/100$ idle error is valid for systems whose gate errors are far from the limits set by coherence times. For example, trapped ion experiments have coherence times $T_2=2T_1$ on the order of $1-10$ seconds with gate times on the order of $10\mu s$. This suggests idle error probabilities of $10^{-5}$ to $10^{-6}$, while two qubit gate infidelities are on the order of $10^{-3}$ to $10^{-4}$ \cite{Trout18,Bermudez18}. On the other hand, the assumption may not hold in systems such as superconducting qubits, whose gates currently achieve infidelities near the coherence limit \cite{Takita17}. Regardless, the concrete schemes we analyze all use the same underlying quantum code family and Clifford gate implementations, so we expect comparisons between them to be robust.
 
In the following sections, all Clifford gates and resource states will be encoded in the Steane code (see \cref{sec:SteaneCode}) which, paired with flag qubits, will allow us to obtain encoded magic states with low overhead. The code distance will be increased through code concatenation.  Since the encoded version of the gates and states are implemented in a fault-tolerant way, the failure probability for each logical fault $E$ of a gate $G$ at a physical error rate $p$ resulting from the malignant event $\text{mal}^{(1)}_{E}$ (where the superscript denotes the level of concatenation) can be upper bounded as

\begin{align}
\text{Pr}[\text{mal}^{(1)}_{E} | G,p] \le \sum^{L_{G}}_{k = \ceil{\frac{d}{2}} }c(k)p^{k} \equiv \Gamma_{G}^{(1)},
\label{eq:Level1LogicFail}
\end{align}
where $c(k)$ denotes the number of weight-$k$ errors which can lead to a logical fault and $L_{G}$ is the total number of circuit locations in the logical gate $G$. At the first concatenation level, we performed Monte-Carlo simulations with the above noise model to estimate the coefficients $c(k)$ for each logical gate and encoded states. 

As was shown in \cite{PR12,CJL16,CJL16b}, \cref{eq:Level1LogicFail} can be generalized to the level-$l$ concatenation level where each physical gate is replaced by its level-$(l-1)$ Rec\footnote{Rectangles (Rec's) are encoded gates with trailing error correction units. Extended rectangles, which are encoded gates with leading and trailing error correction units, are abbreviated as exRec's.} (see \cite{AGP06} for more details). The error rate at the $l$-th concatenation level can be bounded as

\begin{align}
\text{Pr}[\text{mal}^{(l)}_{E} | G,p] \le \sum^{L_{G}}_{k = \ceil{\frac{d}{2}} }c(k)\Big( \Gamma_{G}^{(l-1)} \Big)^{k} \equiv \Gamma_{G}^{(l)}.
\label{eq:LevelLLogicFail}
\end{align}
 
In other words, to estimate the logical failure rate of a level-$l$ gate, the probability of failure for all physical gates $G$ in our level-$l$ simulation will be replaced by $\text{Pr}[\text{mal}^{(l-1)}_{E} | G,p]$ obtained from a level-$(l-1)$ Monte-Carlo simulation. For the physical error rates we consider, higher order terms are found to be relatively small, so we only estimate the leading order terms. More details can be found in \cite{PR12,CJL16,CJL16b}.
 
 \section{Preparing magic states in the Steane code}
\label{sec:SteaneCode}
 
\begin{figure}
	\centering
	\subfloat[\label{fig:HadMeasSimple}]{%
		\includegraphics[width=0.2\textwidth]{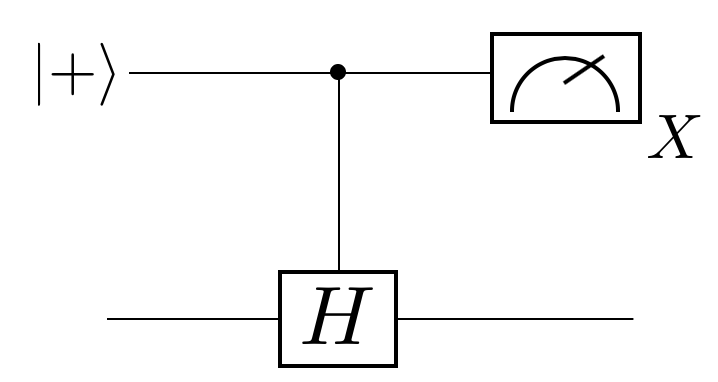}
	}\vfill
	\subfloat[\label{fig:CHdecomp}]{%
		\includegraphics[width=0.3\textwidth]{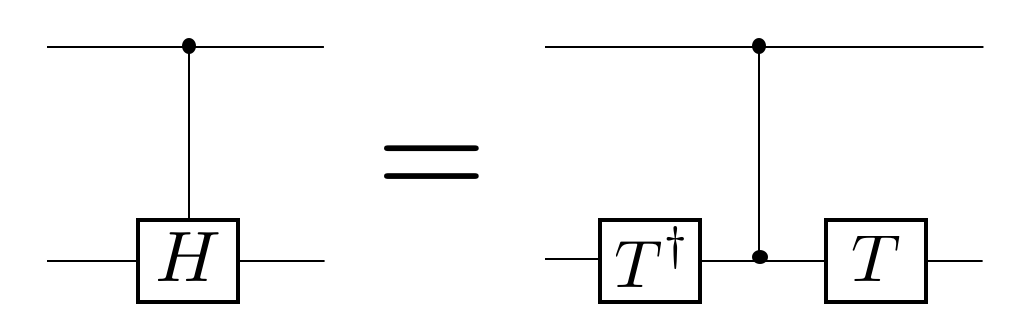}
	}
	\caption{(a) Non-destructive measurement of the Hadamard operator. (b) The $C_{H}$ gate can be decomposed into the gates $T$, $C_{Z}$ and $T^{\dagger}$. As was shown in \cref{fig:TgateCirc}, the $T$ gates can be simulated using only Clifford gates and an $\ket{H}$ resource state. }
	\label{fig:HadMeasUnencoded}
\end{figure}
 
\begin{figure*}
	\centering
	\includegraphics[width=0.95\textwidth]{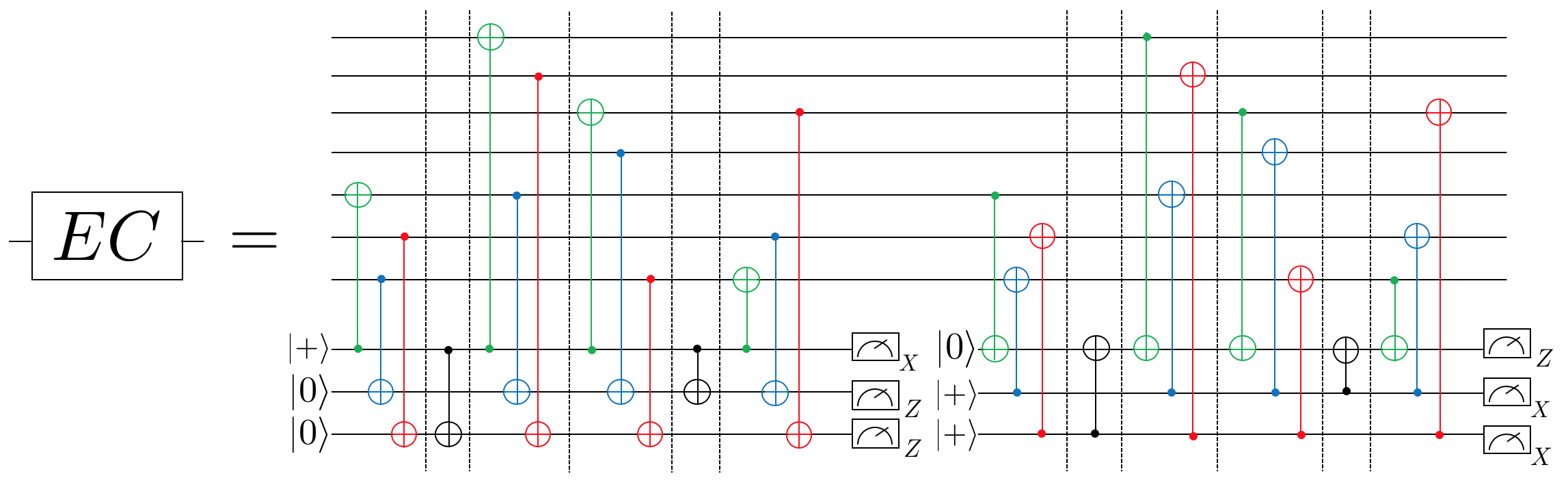}
	\caption{Fault-tolerant error correction circuit introduced by Reichardt \cite{ReichardtFlag18} for measuring the stabilizers of the \codepar{7,1,3} Steane code. The ancilla qubits act as flag qubits that flag if a single fault results in a data-qubit error of weight greater than one.}
	\label{fig:ReichardtECcircuit}
\end{figure*}
 
One way to measure the Hadamard operator using only Clifford gates and an $\ket{H}$ resource state is shown in \cref{fig:HadMeasUnencoded}. To obtain resource states with high-fidelity (which is required for universal quantum computation), one method is to encode them into an error detecting or error correcting code and perform several rounds of state-distillation. If the physical error rate is below some threshold (which depends on the codes and distillation routine), the error rate can be exponentially suppressed with the number of distillation rounds. Another, more direct method, is to fault-tolerantly prepare resource states in a large distance error correcting code. The distance is chosen to obtain the desired logical error suppression. In particular, in the presence of noisy Clifford gates, distillation routines require encoded Clifford operations which can substantially increase the qubit and gate overhead of the routine. We now show how to fault-tolerantly prepare an $\ket{H}$ state using the Steane code and flag-qubits in order to achieve a lower gate and qubit overhead. We will compare the overhead requirements of our methods with that of the Meier-Eastin-Knill (MEK) distillation routine \cite{MEK12}. A review of the MEK distillation routine is given in \cref{sec:MEKdistillationCircuits}.  
 
 \begin{table}
\begin{centering}
\begin{tabular}{c|c}
\codepar{7,1,3} Steane code & \codepar{4,2,2} code \tabularnewline 
\hline
\hline 
$g_1 = XIXIXIX$ & $g_1 = XXXX$ \tabularnewline
\hline 
$g_2 = IIIXXXX$ & $g_2 = ZZZZ$ \tabularnewline
\hline 
$g_3 = IXXIIXX$ & \tabularnewline
\hline
$g_4 = ZIZIZIZ$ & \tabularnewline
\hline 
$g_5 = IIIZZZZ$ & \tabularnewline
\hline 
$g_6 = IZZIIZZ$ & \tabularnewline
\hline
\hline  
$\overline{X} = X^{\otimes 7}$  & $\overline{X}_{1} = XXII$, $\overline{X}_{2} = XIIX$ \tabularnewline
\hline
$\overline{Z} = Z^{\otimes 7}$ & $\overline{Z}_{1} = ZIIZ$, $\overline{Z}_{2} = ZZII$ \tabularnewline
\end{tabular}
\par\end{centering}
\caption{\label{tab:SteaneStabs} Stabilizer generators and logical operators of the \codepar{7,1,3} Steane code and the \codepar{4,2,2} error detecting code.}
\end{table} 
 
 \begin{figure}
	\centering
	\includegraphics[width=0.45\textwidth]{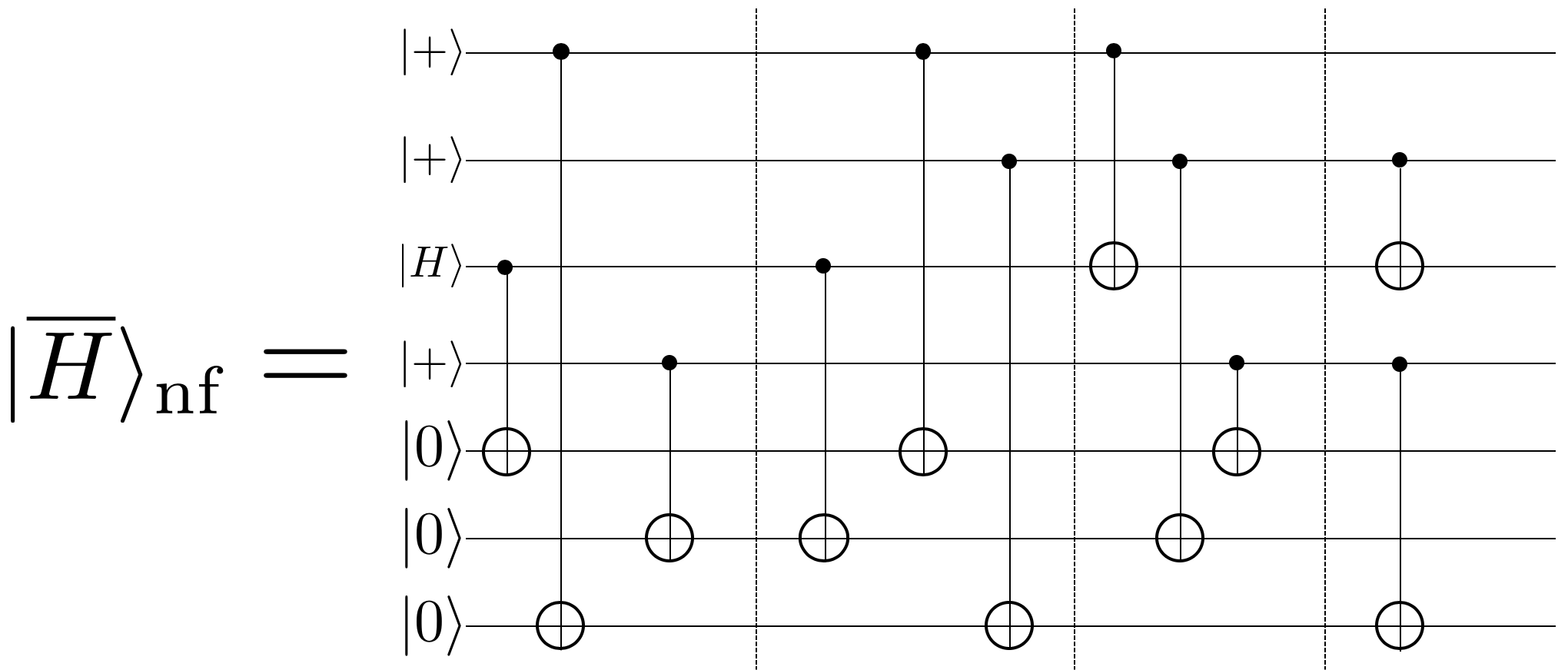}
	\caption{Non-fault-tolerant circuit for preparing a logical $\ket{H}$ state encoded in the \codepar{7,1,3} Steane code. We label this circuit by $\ket{\overline{H}}_{\text{nf}}$. If there are no faults then $\ket{\overline{H}}_{\text{nf}} = \overline{T}|\overline{0}\rangle$.}
	\label{fig:NonFaultTolerantHadCircuit}
\end{figure}
 
 \begin{figure}
	\centering
	\subfloat[\label{fig:HadMeasCircuitDetect}]{%
		\includegraphics[width=0.5\textwidth]{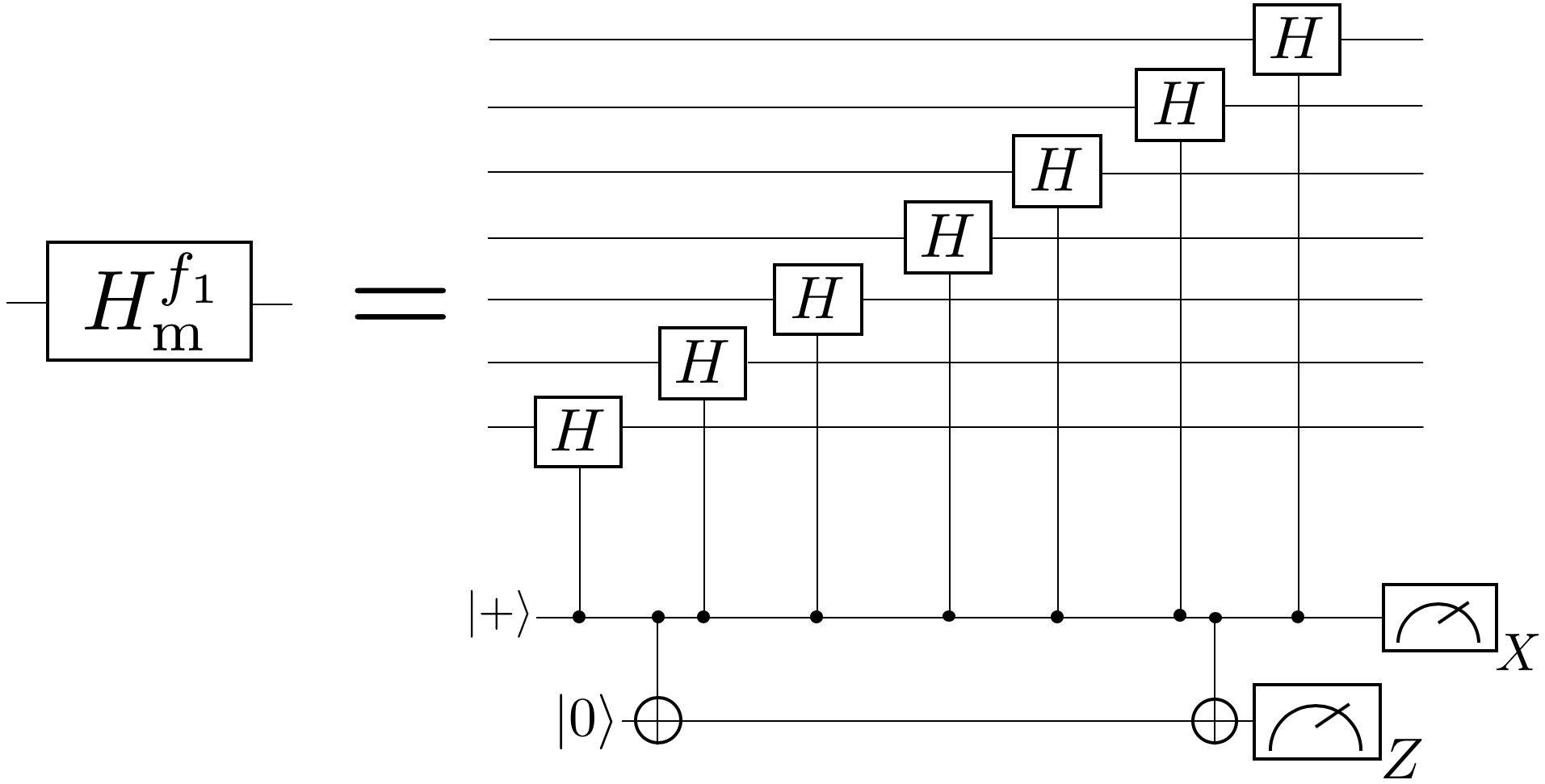}
	}\vfill
	\subfloat[\label{fig:HadMeasDetectScheme}]{%
		\includegraphics[width=0.33\textwidth]{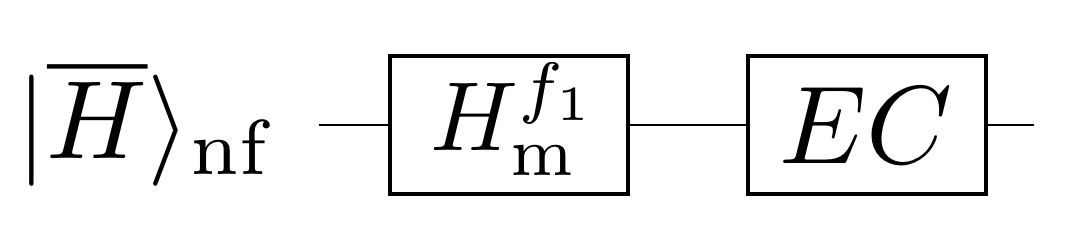}
	}
	\caption{(a) Flag fault-tolerant circuit for measuring the the logical Hadamard operator of the \codepar{7,1,3} Steane code in an error detection scheme. If a single fault results in a data qubit error of weight greater than one, the circuit will flag (the $\ket{0}$ state will be measured as $-1$) and the $\ket{H}$ state preparation scheme will begin anew. Note that the controlled Hadamard gates are implemented as shown in \cref{fig:HadMeasUnencoded}, where the $T$ gates are implemented as in \cref{fig:TgateCirc}. (b) Full fault-tolerant error detection scheme for preparing an encoded $\ket{H}$ state. The EC circuit is given in \cref{fig:ReichardtECcircuit} and the non-fault-tolerant circuit for preparing the encoded $\ket{H}$ state is given in \cref{fig:NonFaultTolerantHadCircuit}.}
	\label{fig:DetectSchemes}
\end{figure}
 
The Steane code is a \codepar{7,1,3} CSS (Calderbank-Shor-Steane) code \cite{CS96,Steane96b} with stabilizer generators and logical operators given in \cref{tab:SteaneStabs}. The Steane code has the property that all Clifford gates can be implemented transversally. In particular, the logical Hadamard operator is given by $\overline{H} = H^{\otimes 7}$. The fault-tolerant preparation of an encoded $\ket{\overline{A_{\frac{\pi}{4}}}}$ (see \cref{eq:Api4State}) state has been previously analyzed in \cite{AGP06}. However, Steane error correction (not to confuse with the Steane code) was used for the EC units and cat states were fault-tolerantly prepared in order to measure $\bar{T}\bar{X}\bar{T}^{\dagger}$ \footnote{Note that the $T$ gates considered in this paper are Clifford-equivalent to the $T$ gates considered in \cite{AGP06}}. The high cost for preparing cat states and implementing Steane EC's results in a large qubit overhead. 

\begin{figure*}
	\centering
	\subfloat[\label{fig:HadMeasCircuitCorrect}]{%
		\includegraphics[width=0.7\textwidth]{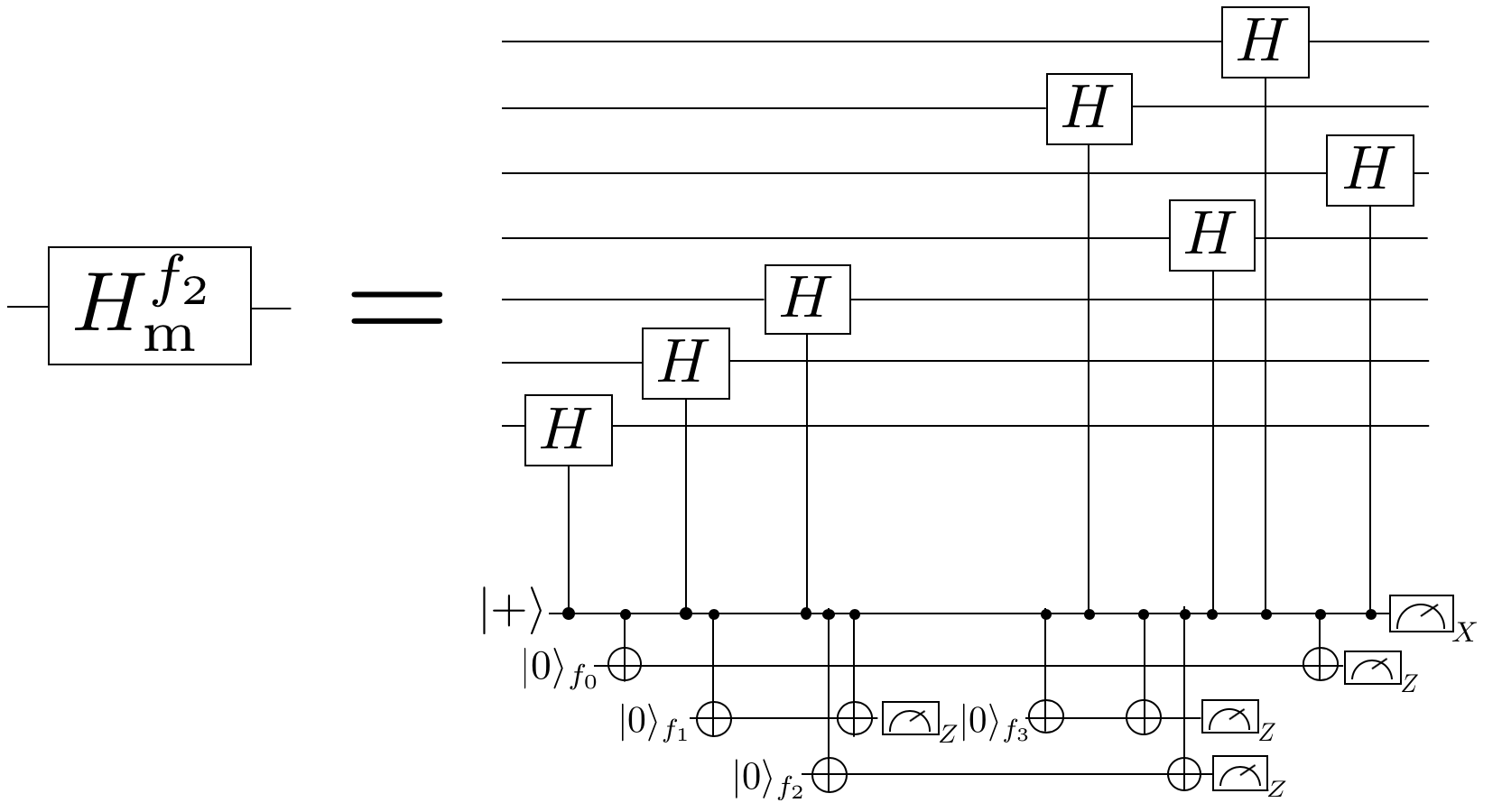}
	}\vfill
	\subfloat[\label{fig:HadMeasCorrectScheme}]{%
		\includegraphics[width=0.5\textwidth]{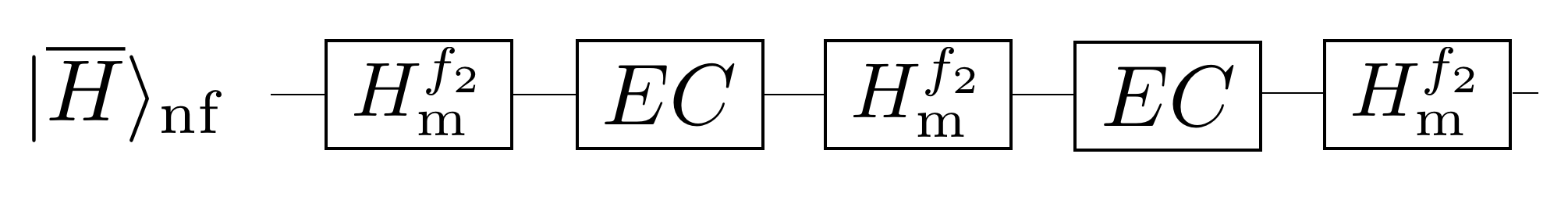}
	}
	\caption{(a) Flag fault-tolerant circuit for measuring the the logical Hadamard operator of the \codepar{7,1,3} Steane code. The extra flag qubits are used to localize errors occurring near the fourth controlled-Hadamard gate since these cannot be distinguished from errors occurring at the other controlled-Hadamard gates. Note that in order to distinguish errors from faults at other locations, the full six bit error syndrome must be considered. That is, we cannot correct $X$ and $Z$ errors separately as is usually done in CSS constructions. (b) Full fault-tolerant error correction scheme for preparing an encoded $\ket{H}$ state. The EC circuit is given in \cref{fig:ReichardtECcircuit} and the non-fault-tolerant circuit for preparing the encoded $\ket{H}$ state is given in \cref{fig:NonFaultTolerantHadCircuit}. If a flag occurs in the third $H_{m}^{f_{2}}$ measurement, an extra round of EC is performed. }
	\label{fig:CorrectSchemes}
\end{figure*}

  \begin{figure*}
	\centering
	\includegraphics[width=0.75\textwidth]{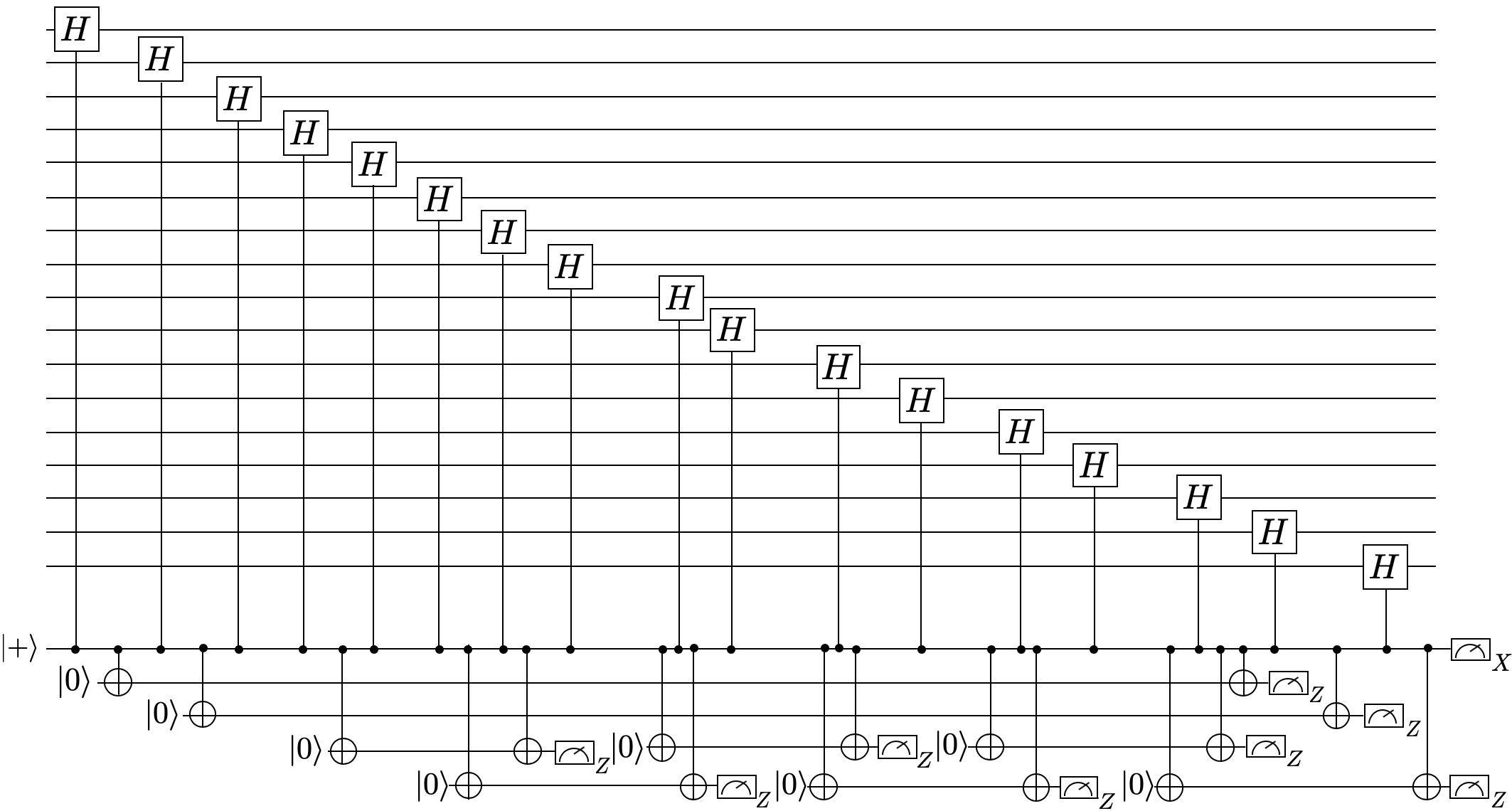}
	\caption{2-flag circuit for measuring the logical Hadamard operator of the \codepar{17,1,5} color code. This circuit can be used in an error detection scheme for fault-tolerantly preparing an $\ket{H}$ state encoded in the \codepar{17,1,5} color code. If there are $v \le 2$ faults which cause a data qubit error of weight greater than $v$, at least one of the flag qubits will flag and the $\ket{H}$ state will be rejected.}
	\label{fig:FlagCircuit17ColorCode}
\end{figure*}

Instead of using Steane-EC circuits, in this work we consider an EC circuit recently introduced by Reichardt \cite{ReichardtFlag18} for measuring the stabilizers of the Steane code with low circuit-depth and which requires only three ancilla qubits (see \cref{fig:ReichardtECcircuit}). The small qubit overhead is achieved with the use of flag qubits which can detect events where errors from a single fault spread to an $X$ or $Z$ error of weight greater than one on the data. The fault-tolerant properties of the circuit are discussed in \cref{subsec:ECcircuit}.
 
Flag-qubits can also be used to fault-tolerantly measure the logical Hadamard operator with only one extra ancilla qubit (and thus do not require the fault-tolerant preparation of cat states). In an error detection scheme, where the $\ket{\overline{H}}$ state is rejected if errors are detected, the circuit in \cref{fig:HadMeasCircuitDetect} can be used to measure the logical Hadamard operator. If a fault results in an $X$ or $Z$ data qubit error of weight greater than one, the flag qubit (ancilla prepared in the $\ket{0}$ state) will be measured as $-1$ and the $\ket{\overline{H}}$ state will be rejected. The full error detecting scheme is shown in \cref{fig:HadMeasDetectScheme} \footnote{In both \cref{fig:HadMeasCircuitDetect} and \cref{fig:HadMeasDetectScheme}, the controlled Hadamard gates are implemented as shown in \cref{fig:HadMeasUnencoded} and for concatenation levels greater than one, the $T$ gates are implemented as in \cref{fig:TgateCirc}. See \cref{app:HprepOverheadAnalysis} for more details.}. Note that our scheme uses only 10-qubits and is shown to be fault-tolerant in \cref{subsec:FTproofErrorDetect}.

Error detection schemes have extra overhead arising from starting the process anew when a state is rejected (although they achieve much higher pseudo-thresholds). Alternatively, in \cref{fig:CorrectSchemes} we show how the $\ket{\overline{H}}$ state can be fault-tolerantly prepared in an error correction scheme using three flag qubits. The details of the implementation and a proof of fault-tolerance is given in \cref{subsec:FTproofErrorCorrect}. Due to the low pseudo-threshold of the error correction scheme, we will focus on the error detection scheme of \cref{fig:DetectSchemes} in the remainder of this manuscript.

The main motivation for concatenating the Steane code in order to increase the code distance is that finding flag circuits to measure high weight operators while maintaining error distinguishability is quite challenging \cite{CB17}. However if a flag circuit satisfying the desired fault-tolerance criteria is found for a small code, the same circuit can be used at level-$l$ where each gate is a level-$(l-1)$ gate. Additionally, if the data used in a quantum computation is encoded in the same code used to prepare the $\ket{H}$ state, the encoded $\ket{H}$ state does not need to be decoded and can be used to directly apply a logical $T$ gate. 

Lastly, we point out that following the 2-flag circuit construction of \cite{CB17}, it is possible to obtain a fault-tolerant circuit to measure the logical Hadamard operator of the \codepar{17,1,5} color code. The circuit is given in \cref{fig:FlagCircuit17ColorCode} and can be used in an error detection scheme analogous to that of \cref{fig:DetectSchemes}. EC circuits for measuring the stabilizers of the \codepar{17,1,5} color code were obtained in \cite{CB17} and require at most four ancilla qubits. One could also measure the stabilizers using standard topological methods at the cost of requiring $\mathcal{O}(n)$ ancillas. This circuit could be useful if a higher distance is required prior to concatenation or if a hybrid state-preparation and magic state distillation scheme is used (see \cref{sec:Discussion} for more details).
 
\section{Meier-Eastin-Knill distillation circuits}
\label{sec:MEKdistillationCircuits}
 
\begin{figure*} 
	\centering
	\subfloat[\label{fig:HMEKcirc1}]{%
		\includegraphics[width=0.95\textwidth]{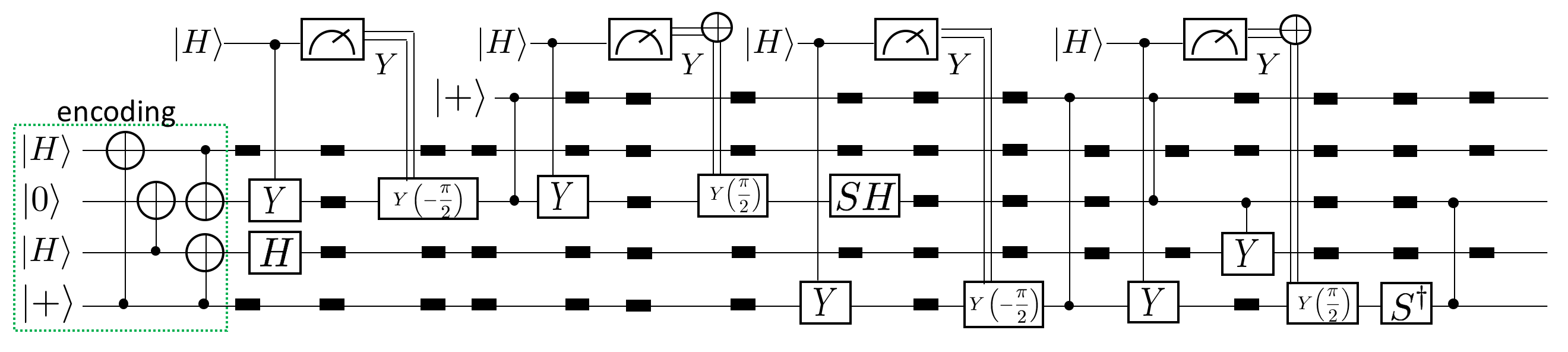}
	}\vfill
	\subfloat[\label{fig:HMEKcirc2}]{%
		\includegraphics[width=0.95\textwidth]{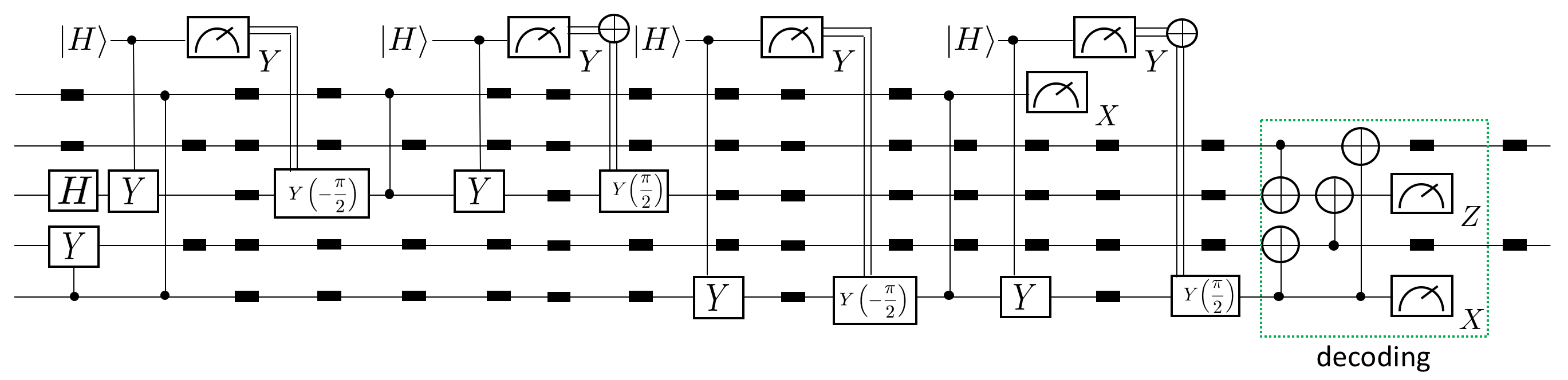}
	}
	\caption{(a) First half of the MEK distillation circuit. (b) Second half of the MEK distillation circuit. The circuit is the same as given in \cite{MEK12} but has been written in detail to illustrate the idle qubit locations (filled rectangles). We reuse the $\ket{H}$ ancilla qubit instead of doing gates in parallel in order to minimize the overhead cost. A total of four $C_{H}$ gates are required to measure the operator $\overline{H}_{1}\overline{H}_{2}$.}
	\label{fig:MEKCircs}
\end{figure*}

In the MEK distillation protocol, $\ket{H}$ states are encoded in the \codepar{4,2,2} error detecting code, whose stabilizer generators and logical operators are given in \cref{tab:SteaneStabs}. For the \codepar{4,2,2} code, the operator $H^{\otimes 4}$ is a valid encoded gate and performs the operation $\overline{H}_{1}\overline{H}_{2}\overline{\text{SWAP}}_{12}$ where $\overline{\text{SWAP}}_{12}$ swaps the two encoded qubits. The circuit in \cref{fig:MEKCircs} performs an encoded measurement of $\overline{H}_{1}\overline{H}_{2}$ on a pair of encoded $\ket{H}$ states and measures the stabilizers of the \codepar{4,2,2} code (note that the circuit applies a Hadamard gate to one of the two encoded qubits). The routine accepts the pair of $\ket{H}$ states if both the $H^{\otimes 4}$ and syndrome measurements are trivial. More details can be found in \cite{MEK12}.

Suppose that the desired output failure probability of the input resource states is $\mathcal{O}(p^{2^{l}})$. Note that a single fault in some of the two-qubit gates in \cref{fig:MEKCircs} can result in multi-qubit errors on the data that go undetected\footnote{This is due to the fact that magic state distillation circuits are fault-tolerant for noise models where errors are only introduced in resource states.}. Since all of the gates in \cref{fig:MEKCircs} fail following the noise model described in \cref{sec:NoiseModelNotation}, they must be encoded in another code (which we choose to be the level-$l$ concatenated Steane code) in order to achieve the desired logical error rate in the output resource states. To provide a fair comparison between the overhead costs of our state preparation scheme with flag qubits and the MEK scheme, all exRec's contain the EC unit of \cref{fig:ReichardtECcircuit}. These features will play an important role when evaluating the qubit and gate overhead of the MEK scheme.

Many magic state distillation protocols use a randomization called twirling \cite{BK05} to diagonalize each magic state in a convenient basis. This greatly simplifies the analysis so that the acceptance probabilities and logical error rates can be found in closed form. However, twirling is not necessary in a physical implementation, since there is always another protocol without the randomization that uses the optimal choice of gates \cite{CB09}. In our case, we observe that the logical error rates increase if we twirl, so we simply use the input states as they occur.
 
\section{Resource overhead comparison}
\label{sec:Discussion}

In this section we compare the overhead cost of our fault-tolerant magic state preparation scheme for preparing an encoded $\ket{H}$ state in the concatenated Steane code to the overhead cost of the MEK scheme for distilling $\ket{H}$ states (also encoded in the concatenated Steane code). In particular, we will compare the qubit and gate overhead cost of both schemes. In what follows, a level-$l$ state or gate will always correspond to its encoded version in $l$ concatenation levels of the Steane code. 

 \begin{figure*} 
\centering
\begin{minipage}{.5\textwidth}
  \centering
  \includegraphics[width=0.98\linewidth]{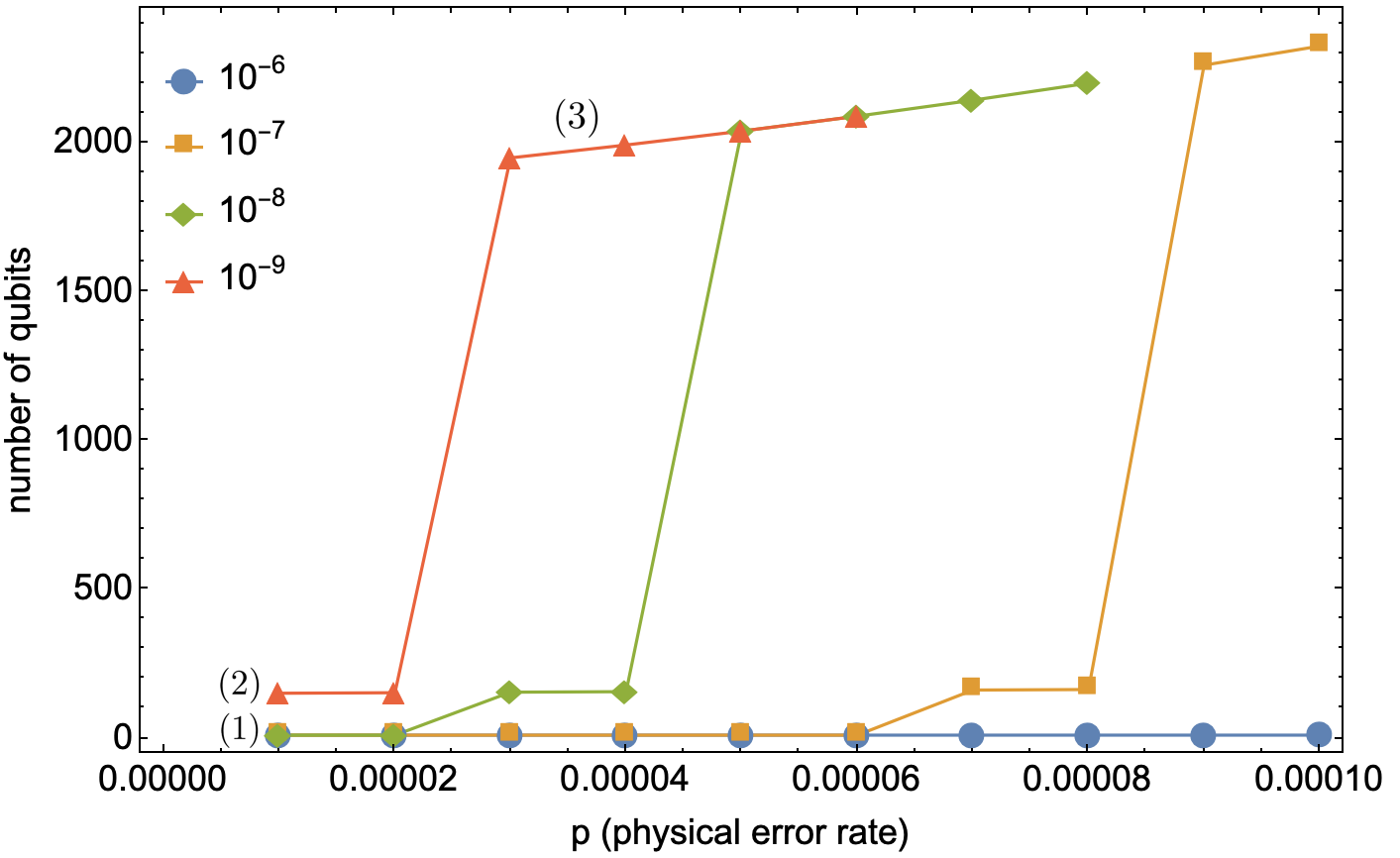}
  \label{fig:CCschemeQubitOverhead1}
\end{minipage}%
\begin{minipage}{.5\textwidth}
  \centering
  \includegraphics[width=0.98\linewidth]{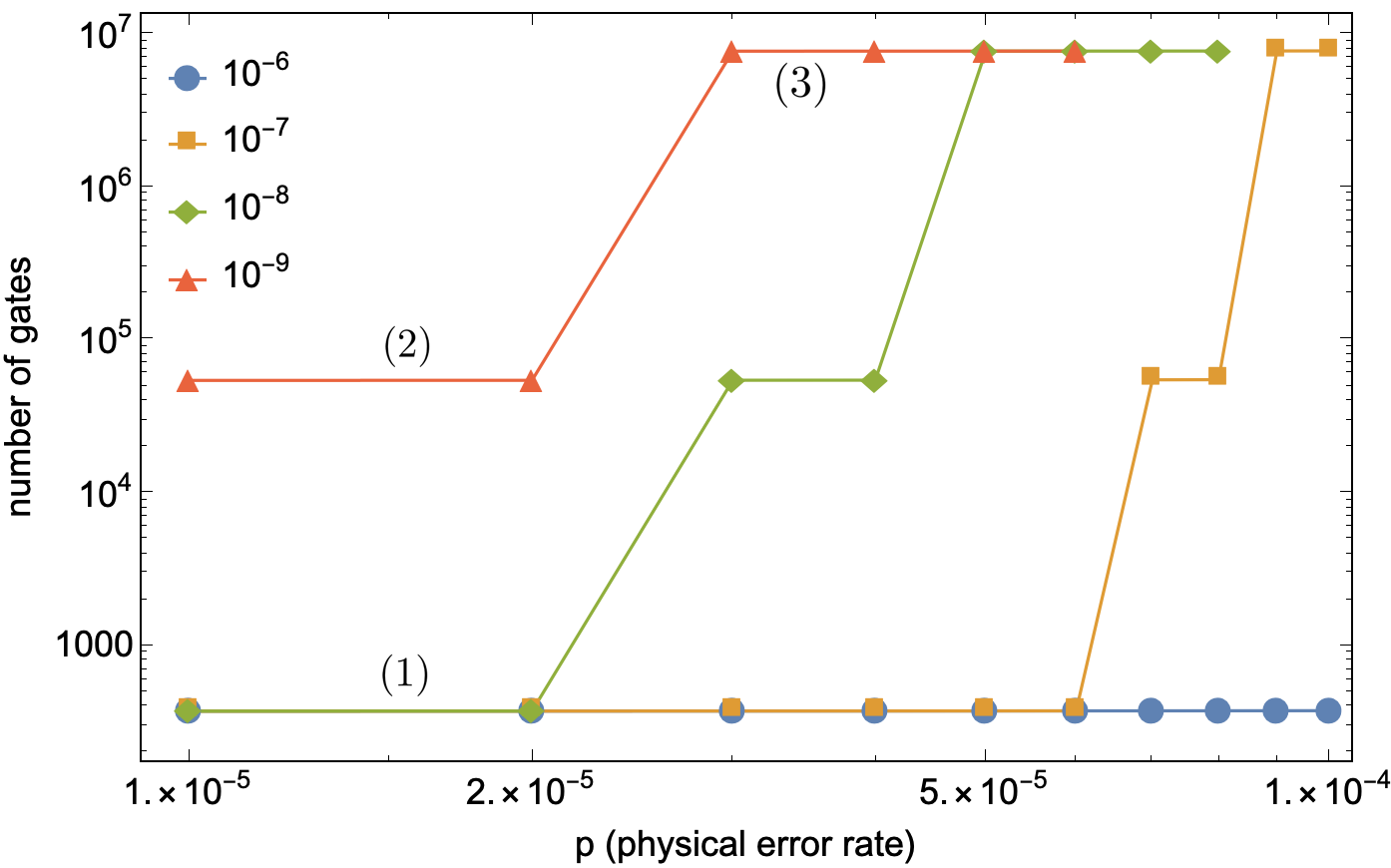}
  \label{fig:CCschemeGateOverhead1}
\end{minipage}
\caption{Qubit and gate overhead for our fault-tolerant magic state preparation scheme using flag qubits. We considering target logical failure rates ranging between $10^{-6}$ to $10^{-9}$. The different jumps in the curves indicate that an additional concatenation level is required in order to achieve the desired logical failure rate. Note that the $y$ axis of the qubit overhead plot is not displayed on a log scale. This is to show the increase in overhead cost due to the probability of rejecting a state when implementing the error detection scheme.}
\label{fig:CCschemePlots}
\end{figure*}

 \begin{figure*} 
\centering
\begin{minipage}{.5\textwidth}
  \centering
  \includegraphics[width=0.98\linewidth]{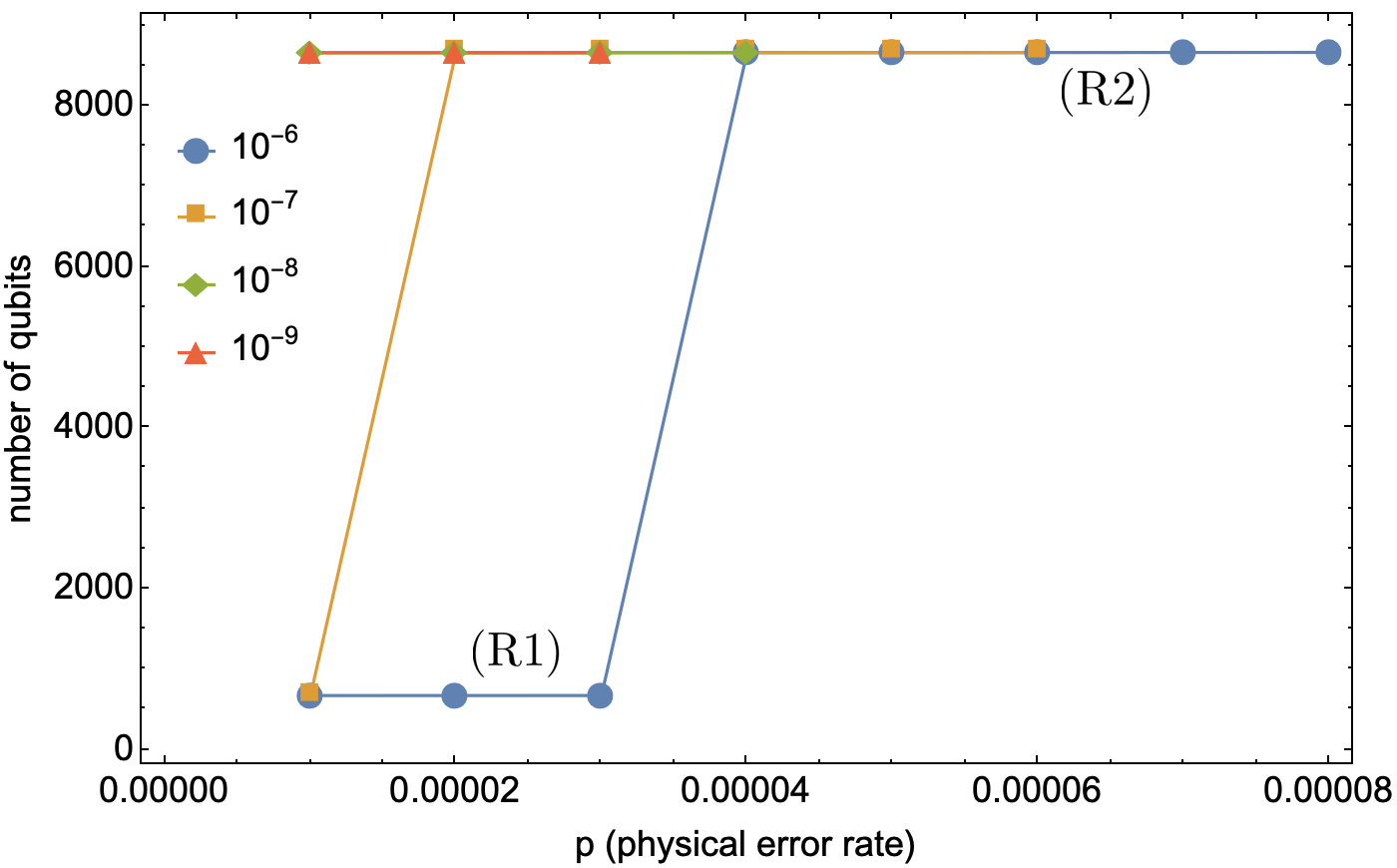}
  \label{fig:MEKschemeQubitOverhead}
\end{minipage}%
\begin{minipage}{.5\textwidth}
  \centering
  \includegraphics[width=0.98\linewidth]{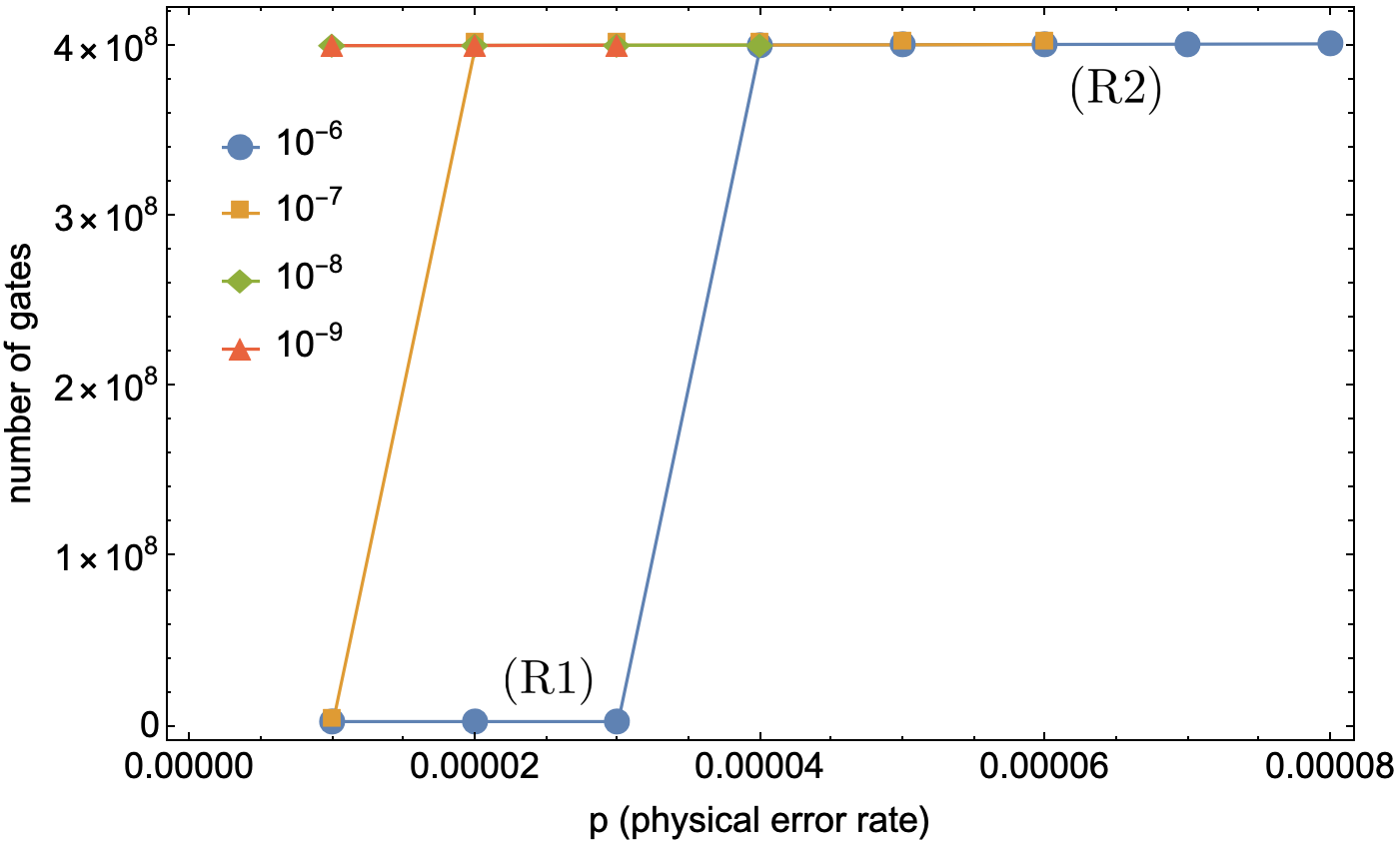}
  \label{fig:MEKschemeGateOverhead}
\end{minipage}
\caption{Qubit and gate overhead for the MEK protocol applied to level-2 and level-3 $\ket{H}$ states. In order to achieve the desired logical failure rate, we teleport a level-$2$ $\ket{H}$ state into a level-$3$ $\ket{H}$ state and apply one round of level-3 MEK. In a level-$l$ simulation, all gates in the MEK circuit are encoded at level-$l$. }
\label{fig:MEKschemePlots}
\end{figure*}

 \begin{figure*} 
\centering
\begin{minipage}{.5\textwidth}
  \centering
  \includegraphics[width=0.98\linewidth]{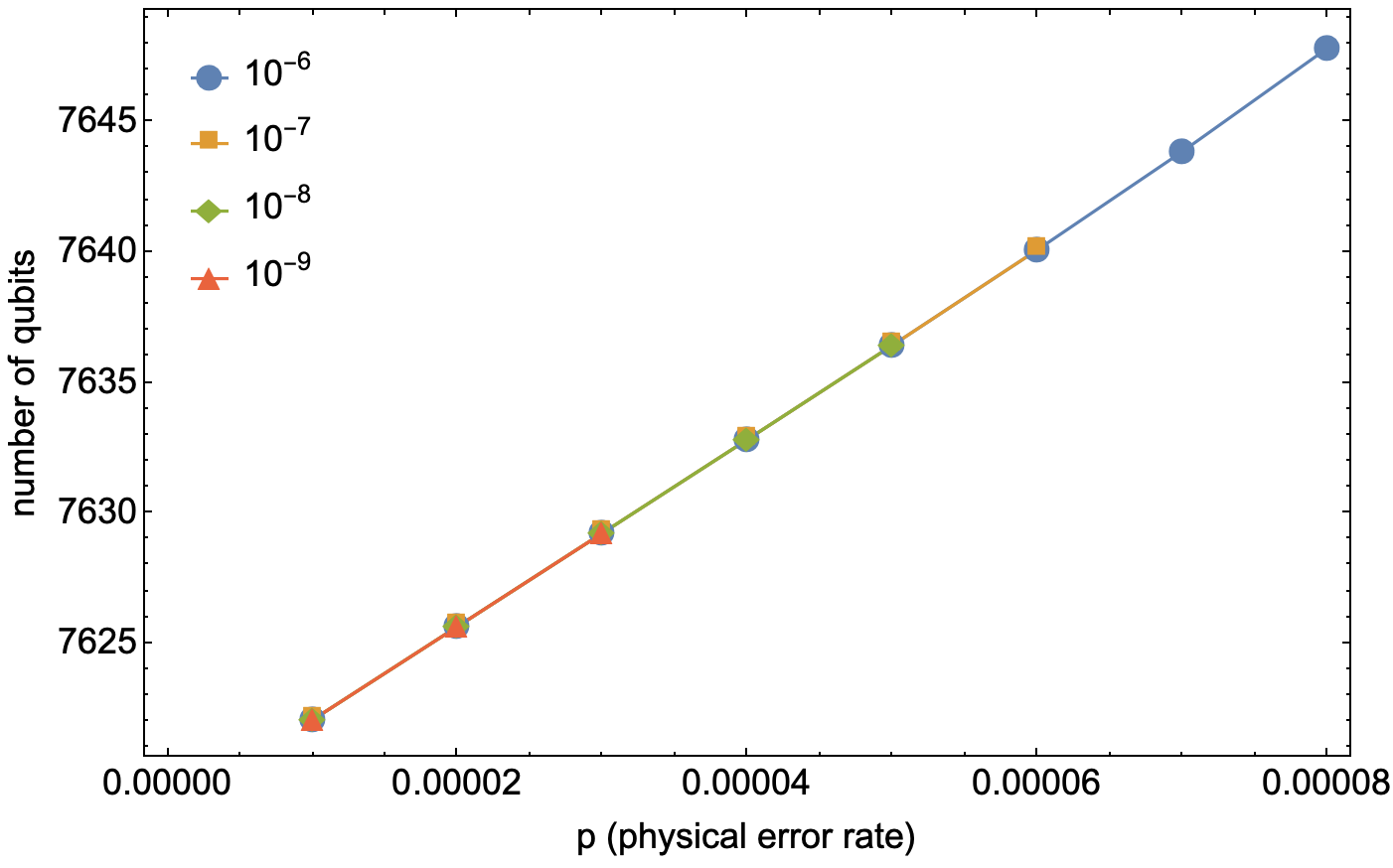}
  \label{fig:MEKCCqubitOverhead}
\end{minipage}%
\begin{minipage}{.5\textwidth}
  \centering
  \includegraphics[width=0.98\linewidth]{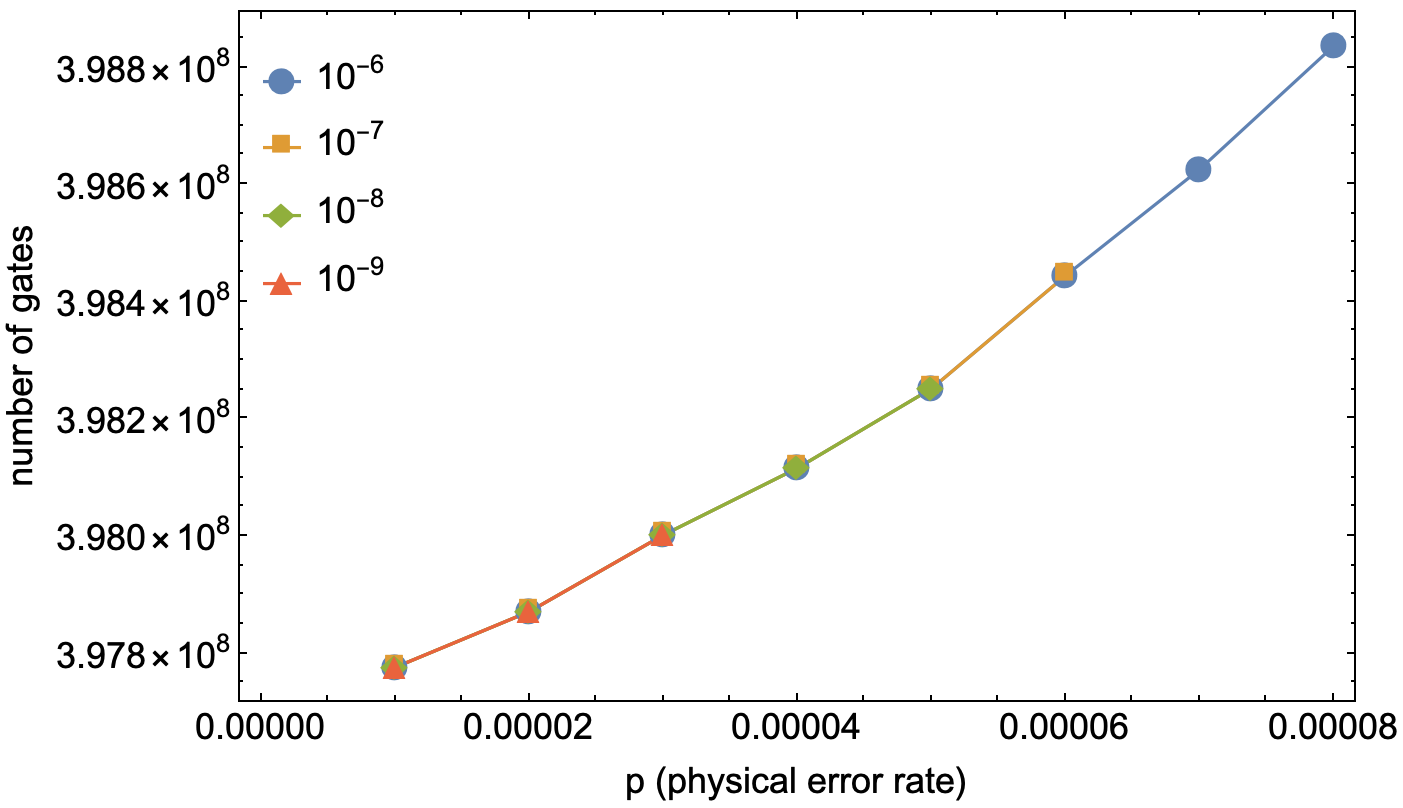}
  \label{fig:MEKCCGateOverhead}
\end{minipage}
\caption{Qubit and gate overhead for the hybrid scheme. First a level-$2$ $\ket{H}$ state is fault-tolerantly prepared using the methods of \cref{fig:DetectSchemes}. Afterwards the state is teleported to a level-$3$ state and a round of level-$3$ MEK is performed.}
\label{fig:MEKCCschemePlots}
\end{figure*}

For the MEK magic state distillation protocol, we consider two different approaches. In the first approach, physical $\ket{H}$ states are teleported into level-2 $\ket{H}$ states and a subsequent round of MEK (with level-2 Clifford gates) is applied to produce a state with a logical error rate well-approximated by the form $ap^{2} + bp^{4}$ below the level-2 pseudo-threshold. The $\mathcal{O}(p^2)$ term is dominant at low error rates, and the $\mathcal{O}(p^4)$ term is dominant at error rates near the pseudo-threshold. To obtain further error suppression, we then teleport the distilled state into a level-3 state and perform a level-3 round of MEK to obtain a state with a logical error rate of the form $a'p^{4} + b'p^{8}$. An additional round of MEK could be performed to obtain an $\ket{H}$ state with logical error rate $\mathcal{O}(p^{8})$. However, for the physical error rates considered in this work ($p \in [ 10^{-5}, 10^{-4} ]$), we found that there is no advantage in doing so.

In the second approach, we consider a hybrid scheme where a level-2 $\ket{H}$ state is first prepared using our fault-tolerant flag preparation scheme. The state is then teleported into a level-3 state and a round of MEK is performed resulting in an $\ket{H}$ state with logical error rate $\mathcal{O}(p^{8})$. All teleportation schemes are performed using the methods described in \cref{app:GateTeleportation}.

The qubit and gate overhead results for the magic state preparation scheme with flag qubits and the MEK schemes are shown in \cref{fig:CCschemePlots,fig:MEKschemePlots,fig:MEKCCschemePlots}. Details of the analysis leading to the plots are given in \cref{app:HprepOverheadAnalysis,app:MEKOverheadAnalysis}.

Comparing the results, it is clear that the qubit and gate overhead cost of the fault-tolerant $\ket{H}$ state preparation scheme of \cref{fig:DetectSchemes} is substantially smaller than schemes involving MEK. The primary reasons for this are due to the high pseudo-threshold of the error detection scheme as well as the small size of the circuits compared to the circuits used by MEK (where we used the teleportation scheme of \cref{app:GateTeleportation} to inject states at a higher concatenation level). Further, note that the hybrid scheme (\cref{fig:MEKCCschemePlots}) has smaller qubit and gate overhead costs compared to the full MEK scheme (\cref{fig:MEKschemePlots}), although the overhead is still much larger than the error detection scheme using flag qubits. 

Lastly, we point out that due to the lower logical failure rates of the error detection scheme using flag qubits, three concatenation levels were sufficient to achieve logical failure rates of $10^{-9}$ for larger physical error rates compared to MEK (the same is true for other target logical failure rates). For example, suppose one wants to achieve a logical failure rate $p_{L} \le 10^{-9}$. In our error detection scheme, one must use at least four concatenation levels for physical error rates $p > 6 \times 10^{-5}$. For the MEK scheme, one requires four concatenation levels for physical error rates $p > 4\times 10^{-5}$. Hence for physical error rates $4\times 10^{-5} \le p \le 6 \times 10^{-5}$, the error detection scheme will require fewer qubits by \textit{several orders of magnitude} compared to MEK (see \cref{fig:CCschemePlots,fig:MEKschemePlots}). 

\section{Conclusion}
\label{sec:conclusion}

The high cost of universal fault-tolerant logic is a challenging and enduring problem. In this work, we propose an alternative to magic state distillation that is based on recently discovered flag fault-tolerance techniques. We have designed a flag-based magic state preparation scheme that reduces the number of qubits and operations to prepare a reliable magic state, and does so by orders of magnitude in some error regimes. Furthermore, the error-detection-based state preparation circuit at level-1 has a high pseudothreshold and uses a total of 10 qubits. The circuit accepts around $80\%$ of the time and has logical error rates near $10^{-4}$ at physical error rates near $3\times 10^{-3}$ (see \cref{tab:PrepSimulationResults}), making it an interesting candidate for experimental consideration.

The magic state preparation scheme relies on a transversal Hadamard operator and a flag circuit for measuring it fault-tolerantly. So that the operator's weight does not grow, we concatenate the code with itself to achieve some low logical error rate. The threshold is ultimately that of the concatenated Steane code in all of the schemes we analyze. It would be interesting to account for more realistic constraints on qubit connectivity, and to consider how to extend these techniques to families of higher distance (non-concatenated) codes, but we leave this to future work.

Another interesting application of our work is its use in state injection schemes \cite{LR14,LiMagicState17}. Using the fault-tolerant circuits for distance-three and five color codes in \cref{fig:DetectSchemes,fig:FlagCircuit17ColorCode} (the $w$-flag circuits described in \cite{CB17} could also be used for higher distance color codes), encoded magic states with logical failure rates of $\mathcal{O}(p^2)$ and $\mathcal{O}(p^3)$ can be produced. Lattice surgery techniques could then be performed to further distill the input magic states \cite{LR14,VuilloteLatticeSurgery19}. We note that the state injection scheme in \cite{LiMagicState17} produces encoded magic states with logical failure rate $\mathcal{O}(p)$ and thus for low enough physical error rates, our approach could achieve lower logical failure rates.

If codes such as topological codes were used to perform the MEK protocol at high physical error rates, then magic state distillation might outperform the flag-based scheme, since a large number of concatenation levels would be necessary. However, the ideas we have presented may be broadly applicable to improve the pseudothreshold of magic state preparation circuits using codes other than the Steane code. In this direction, we have given a 2-flag circuit for fault-tolerantly measuring logical Hadamard of the 17-qubit color code. If lower logical error rates are needed, states could then be used in a hybrid approach that takes advantage of the topological code threshold. Lastly, one could use the $w$-flag circuit construction and EC circuits of \cite{CB17} to fault-tolerantly prepare magic states for the family of color codes on a hexagonal lattice.

\section{Acknowledgements}
C.C. acknowledges IBM for its hospitality where all of this work was completed. C.C. also acknowledges the support of NSERC through the PGS D scholarship. We thank Sergey Bravyi, Earl Campbell, Tomas Jochym-O'Connor, and Ted Yoder for useful discussions. We thank Earl Campbell for sharing observations about the MEK protocol.

\newpage 

\bibliographystyle{unsrtnat} 
\bibliography{bibtex_chamberland}

\clearpage
\appendix

\section{Proof of fault-tolerance for the magic state preparation schemes}
\label{app:FaultToleranceProofs}

In this appendix we provide proofs showing that our magic state preparation schemes for the $\ket{H}$ state are fault-tolerant. In what follows, a state-preparation scheme will be called fault-tolerant if the following two conditions are satisfied \cite{Gottesman2010}

\begin{definition}{\underline{Fault-tolerant state preparation}}
	
	For $t = \lfloor (d-1)/2\rfloor$, a state-preparation protocol using a distance-$d$ stabilizer code $C$ is $t$-fault-tolerant if the following two conditions are satisfied:
	\begin{enumerate}
		\item If there are $s$ faults during the state-preparation protocol with $s \le t$, the resulting state differs from a codeword by an error of at most weight $s$.
		\item If there are $s$ faults during the state-preparation protocol with $s \le t$, then ideally decoding the output state results in the same state that would be obtained from the fault-free state-preparation scheme. 
	\end{enumerate}
	\label{Def:FaultTolerantPrep}
\end{definition}

In \cref{Def:FaultTolerantPrep}, ideally decoding is equivalent to performing fault-free error correction. Now suppose $\ket{\overline{\psi}}$ is the encoded state to be prepared. If there are $s \le t$ faults during a state-preparation protocol satisfying the criteria in \cref{Def:FaultTolerantPrep}, then the output state will have the form $E \ket{\overline{\psi}}$ with $\text{wt}(E) \le s$ (the output state will encode the correct state with no more than $s$ errors).

For CSS codes, this definition can be applied independently to $X$ and $Z$ errors.

\subsection{Error correction circuit}
\label{subsec:ECcircuit}

In this section we provide more details on the properties of the EC circuit of \cref{fig:ReichardtECcircuit} which is used in all of our schemes. 

The first half of the circuit measures the $XIXIXIX$ (green CNOT's), $IIIZZZZ$ (blue CNOT's) and $IZZIIZZ$ (red CNOT's) stabilizers of the Steane code. The second half of the circuit measures the remaining stabilizers. Given that the CNOT gates are not transversal, it is possible for a single fault to result in a weight-two $X$ or $Z$ data-qubit error. However the ancilla qubits also act as flag qubits which can be used to detect such events. 

Let us assume that a single-fault occurs. For the first half of the EC circuit, the possible weight-two errors that can arise from a single fault are $Z_{4}Z_{6}$, $Z_{3}Z_{7}$, and $X_{3}X_{7}$. In the case of $Z_{4}Z_{6}$ and $Z_{3}Z_{7}$, the measurement in the $X$-basis will flag (with the two $Z$-basis measurements being $+1$), at which point the entire syndrome measurement is repeated. Since $Z_{4}Z_{6}$ and $Z_{3}Z_{7}$ are errors that have different syndromes than all other errors arising from a single fault leading to a $-++$ measurement outcome of the first three ancilla qubits, after the syndrome measurement is repeated these errors can be distinguished and corrected.

The same argument applies to the case where a single fault leads to the $X_{3}X_{7}$ data-qubit error. But this time the $X_{4}$ error has the same syndrome as $X_{3}X_{7}$ ($s(X_{4}) = s(X_{3}X_{7}) = 010000$). However, the faults causing an $X_{3}X_{4}$ error result in a $+--$ measurement outcome of the first three ancillas whereas $X_{4}$ results in $+-+$. Thus when the syndrome measurement is repeated and the flag outcomes are taken into account, the errors can be distinguished and corrected. 

An analogous analysis can be applied to the second half of the EC circuit. See \cite{ReichardtFlag18} for more details. 

\subsection{Fault-tolerance proof for the error detection scheme}
\label{subsec:FTproofErrorDetect}

Recall that the error detection scheme can be decomposed into three components as shown in \cref{fig:HadMeasDetectScheme}. Since $t=1$ for the \codepar{7,1,3} Steane code, we must show that if a single fault occurs in any of the three components, both criteria in \cref{Def:FaultTolerantPrep} will be satisfied. 

\underline{Case 1: fault in $\ket{\overline{H}}_{\text{nf}}$ (see \cref{fig:NonFaultTolerantHadCircuit})}. Since the Steane code is a perfect CSS code and the $\ket{\overline{H}}_{\text{nf}}$ circuit is not fault-tolerant, the output state of the first component will have the form 

\begin{align}
\ket{\psi^{(1)}} = E^{(x)}_{i}E^{(z)}_{j} \bar{E}^{(x)}_{i}\bar{E}^{(z)}_{j} \ket{\overline{H}},
\label{eq:OutHadNonFT1}
\end{align}
where $E^{(x)}_{i} \in \{I, X_{i} \}$ and $E^{(z)}_{j} \in \{I, Z_{j} \}$ are single-qubit errors on qubits $i$ and $j$ and $\bar{E}^{(x)}_{i} \in \{ I, \overline{X} \}$, $\bar{E}^{(z)}_{j} \in \{ I, \overline{Z} \}$ are logical operators of the Steane code.

Now the output state of the second component of \cref{fig:HadMeasDetectScheme}, including the contribution from the first ancilla qubit (which will subsequently be measured in the $X$-basis), is given by
\begin{widetext}
\begin{equation}
\ket{\psi^{(2)}}  = \frac{1}{2} \Big\{ \Big(E_{i}^{(x)}E_{j}^{(z)}\bar{E}_{i}^{(x)}\bar{E}_{j}^{(z)} + E_{i}^{(z)}E_{j}^{(x)}\bar{E}_{i}^{(z)}\bar{E}_{j}^{(x)} \Big) \ket{\overline{H}}\ket{+} +  \Big(E_{i}^{(x)}E_{j}^{(z)}\bar{E}_{i}^{(x)}\bar{E}_{j}^{(z)} - E_{i}^{(z)}E_{j}^{(x)}\bar{E}_{i}^{(z)}\bar{E}_{j}^{(x)} \Big) \ket{\overline{H}} \ket{-}\Big\}.
\label{eq:OutHadNonFT2}
\end{equation}
\end{widetext}
If $E_{i}^{(x)} \neq I$ or $E_{j}^{(z)} \neq I$, then the single-qubit error will be detected by the subsequent EC and the state will be rejected. Thus let us assume that $E_{i}^{(x)} = E_{j}^{(z)} =I$ so that 
\begin{align}
&\ket{\psi^{(2)}} = \frac{1}{2} \Big\{  \Big(\bar{E}_{i}^{(x)}\bar{E}_{j}^{(z)} + \bar{E}_{i}^{(z)}\bar{E}_{j}^{(x)} \Big) \ket{\overline{H}} \ket{+} \nonumber \\ 
&+  \Big(\bar{E}_{i}^{(x)}\bar{E}_{j}^{(z)} - \bar{E}_{i}^{(z)}\bar{E}_{j}^{(x)} \Big) \ket{\overline{H}} \ket{-}\Big\}.
\end{align}

If $\bar{E}_{i}^{(x)}\bar{E}_{j}^{(z)} = \overline{Y}$, then $ \ket{\psi^{(2)}} = \ket{-\overline{H}}\ket{-}$ and the ancilla measurement will be $-1$ resulting in rejection. Thus suppose that $\bar{E}_{i}^{(x)}\bar{E}_{j}^{(z)}$ is $\overline{X}$ or $\overline{Z}$ so that 
\begin{align}
\ket{\psi^{(2)}} = \frac{1}{\sqrt{2}} \Big\{  \ket{\overline{H}}\ket{+} \mp i \ket{-\overline{H}}\ket{-} \Big\},
\label{eq:BeforeCollapse1}
\end{align}
where we used the identities $H = \frac{1}{\sqrt{2}} (X+Z)$, $XHX = \frac{1}{\sqrt{2}} (X-Z)$ and $ZHZ = \frac{1}{\sqrt{2}} (Z-X)$. From \cref{eq:BeforeCollapse1}, we see that $\ket{\psi^{(2)}}$ will be accepted with probability $1/2$ resulting in the state $ \ket{\overline{H}}$. Thus if a single fault occurs in the first component of \cref{fig:HadMeasDetectScheme}, an accepted state will be $\ket{\overline{H}}$.

\underline{Case 2: fault in $H_{m}^{f_{1}}$ (see \cref{fig:HadMeasCircuitDetect})}. Since there are no faults in $\ket{\overline{H}}_{\text{nf}}$, the input to the circuit will be $\ket{\overline{H}}$. If the fault results in any measurement outcome to be $-1$, the state will be rejected. Thus we only consider faults such that the measurement outcome of all ancilla's is $+1$. Since the circuit will flag if a single fault results in a data-qubit error of weight greater than one, the output state of the circuit will be $\ket{\psi^{(2)}_{\text{out}}} = E_i^{(x)}E_{j}^{(z)}\ket{\overline{H}}$ with $\text{wt}(E_i^{(x)}) \le 1$ and $\text{wt}(E_j^{(z)}) \le 1$. If $E_i^{(x)}E_{j}^{(z)}$ is non-trivial, it will be detected by the subsequent EC circuit and the state will be rejected. Hence an accepted state will be $\ket{\overline{H}}$.

\underline{Case 3: fault in the EC circuit (see \cref{fig:ReichardtECcircuit})}. Since there are no faults in the first two components of \cref{fig:HadMeasDetectScheme}, the input state to the EC will be $\ket{\psi^{(3)}} = \ket{\overline{H}}$. If a fault causes a non-trivial measurement, the state will be rejected. Hence let us consider the case where a single fault results in an error which goes undetected. A single fault in the EC can result in a data-qubit error of the form $E_{i}^{(x)}E_{j}^{(z)}$ where $\text{wt}(E_{i}^{(x)}) \le 1$ and $\text{wt}(E_{j}^{(z)}) \le 1$ without any flags\footnote{As an example, consider the error $Z\otimes X$ on the fourth black CNOT of \cref{fig:ReichardtECcircuit} which results in the data-qubit error $X_6Z_7$.}. However, due to the fault-tolerant properties of the EC, a fault resulting in a data error with $\text{wt}(E_{i}^{(x)}) \ge 2$ or $\text{wt}(E_{j}^{(z)}) \ge 2$ will cause one of the flag qubits to flag. Hence if the state is accepted (all measurements are trivial), the output state of the EC will have the form

\begin{figure}
	\centering
	\subfloat[\label{fig:XpropCH}]{%
		\includegraphics[width=0.4\textwidth]{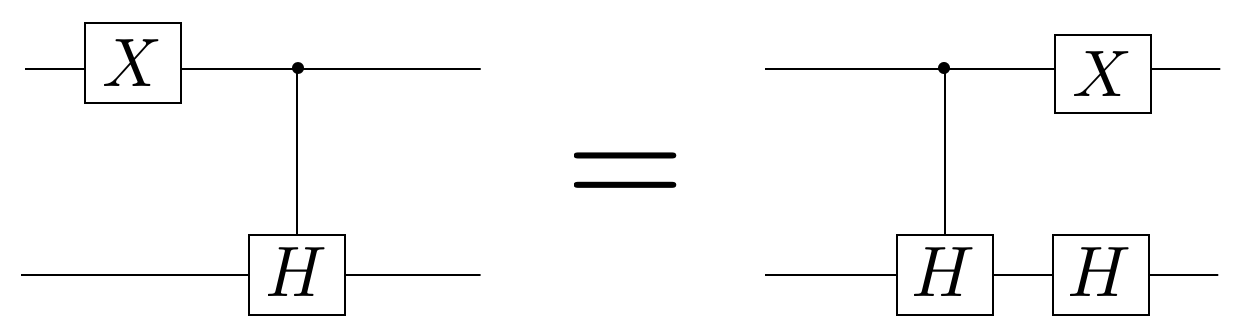}
	}\vfill
	\subfloat[\label{fig:ZpropCH}]{%
		\includegraphics[width=0.4\textwidth]{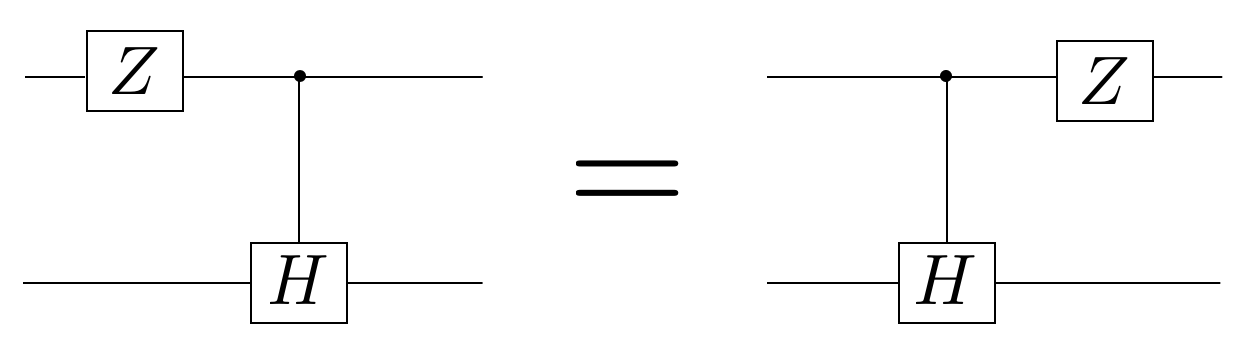}
	}
	\caption{(a) Propagation of $X$ errors through the control-qubit of a controlled-Hadamard gate. This induces a Hadamard error on the target qubit. (b) Propagation of a $Z$ error through the control-qubit of a controlled-Hadamard gate.}
	\label{fig:PropErrorsCH}
\end{figure}

\begin{align}
\ket{\psi_{\text{out}}} = E_{i}^{(x)}E_{j}^{(z)} \ket{\overline{H}},
\end{align}
with $\text{wt}(E_{i}^{(x)}) \le 1$ and $\text{wt}(E_{j}^{(z)}) \le 1$. Since errors of the form $E=X_{i}Z_{j}$ are correctable by the Steane code, fault-free error correction of an accepted state of the protocol in \cref{fig:HadMeasDetectScheme} will always result in the state $\ket{\overline{H}}$. We conclude that if there is at most one fault, both criteria in \cref{Def:FaultTolerantPrep} will be satisfied. 

\subsection{Fault-tolerance proof for the error correction scheme}
\label{subsec:FTproofErrorCorrect}

\subsubsection{Error distinguishability}
\label{subsec:ErrorDistinguishability}

Recall that for the error correction scheme, the circuit for measuring the logical Hadamard operator is given in \cref{fig:HadMeasCircuitCorrect}. The first thing to notice is that the controlled-Hadamard gates are implemented in a different order compared to the gates in \cref{fig:HadMeasCircuitDetect}. If a fault occurs at one of the controlled-Hadamard gates resulting in an $X$ or $Y$ error on the control-qubit, the resulting data qubit error can be expressed as a product of Hadamard errors and Pauli errors (see \cref{fig:PropErrorsCH}). For the second to sixth controlled-Hadamard gates, all errors arising from a single fault at one of these locations which propagate to the data qubits (causing at least one of the flag qubits to flag) must be distinguishable (since $\overline{H} = H^{\otimes 7}$, a fault occurring at the first controlled-Hadamard gate will result in an $X$ and $Z$ data-qubit error of weight at most one). This is only possible if errors are corrected based on their full syndrome\footnote{Instead of using a three-bit syndrome for correcting $X$ errors and separately using the other three-bit syndrome to correct $Z$ errors, one would use the full six-bit syndrome to correct the data errors. Note that we only correct using the full six-bit syndrome if there are flags. Otherwise, we perform the standard CSS correction of $X$ and $Z$ errors separately.} and with a carefully chosen ordering of the controlled-Hadamard gates. Performing a numerical search of all $7! = 5040$ permutations of the controlled-Hadamard gates, there was no permutation that allowed all possible errors (arising from a single fault) to be distinguished. However, ignoring a fault on the fourth controlled-Hadamard gate, a numerical search showed that the ordering found in \cref{fig:HadMeasCircuitCorrect} allows errors arising from a single fault to be distinguished. Consequently, we need to have the ability to isolate errors arising from a fault at the fourth controlled-Hadamard location. This can be achieved using the two extra flag qubits shown in \cref{fig:HadMeasCircuitCorrect} ($\ket{0}_{f_{1}}$ and $\ket{0}_{f_{3}}$ can be measured using the same qubit). 

Suppose for example that an $X$ error arises on the control-qubit of the fourth controlled-Hadamard gate resulting in the data qubit error $P_{2}H_{1}H_{3}H_{4}$, where $P_{2} \in \{ I, X_{2}, Y_{2}, Z_{2} \}$. Then $\ket{0}_{f_{0}}$, $\ket{0}_{f_{2}}$ and $\ket{0}_{f_{3}}$ will flag. We could then apply $H_{1}H_{3}H_{4}$ to the data qubit and be left with the single-qubit Pauli $P_{2}$. Of course, if a fault occurs on the first CNOT connecting $\ket{0}_{f_{3}}$ to the $\ket{+}$ state resulting in an $X \otimes X$ error, $\ket{0}_{f_{3}}$ will not flag but $\ket{0}_{f_{2}}$ will flag (in addition to $\ket{0}_{f_{0}}$). After applying $H_{1}H_{3}H_{4}$, the resulting data qubit error would be $H_{2}$. Going through all possible cases of a single fault at one of the CNOT gates interacting the ancillas $\ket{0}_{f_{1}}$, $\ket{0}_{f_{2}}$ and $\ket{0}_{f_{3}}$ with the $\ket{+}$ state, we can guarantee error distinguishability by applying $H_{1}H_{3}H_{4}$ to the data if the following combinations of flag qubits flag:
\begin{enumerate}
	\item $\ket{0}_{f_{0}}$ and  $\ket{0}_{f_{2}}$ flag.
	\item $\ket{0}_{f_{0}}$ and  $\ket{0}_{f_{3}}$ flag.
	\item $\ket{0}_{f_{0}}$, $\ket{0}_{f_{2}}$ and $\ket{0}_{f_{3}}$ flag.	
\end{enumerate}
Note that $\ket{0}_{f_{0}}$ would not flag if a single measurement error of the flag qubits $\ket{0}_{f_{1}}$, $\ket{0}_{f_{2}}$ or $\ket{0}_{f_{3}}$ occurred.

When following the circuit in \cref{fig:HadMeasCorrectScheme} by an EC circuit, we will use a lookup-table containing all errors arising from a single fault resulting in a flag (which are all distinguishable) in order to correct.

\subsubsection{Fault-tolerance proof}
\label{eq:FTerrorCorrProof}

Given the above for applying Hadamard corrections to the data qubits depending on the flag outcomes (thus guaranteeing error distinguishability if a single fault occurs), we now show that our error correcting scheme shown in \cref{fig:HadMeasCorrectScheme} satisfies both criteria in \cref{Def:FaultTolerantPrep}. In what follows, $\ket{\psi_{\text{final}}}$ will correspond to the output state of the circuit in \cref{fig:HadMeasCorrectScheme}. Also, the state at the output of the $i$'th circuit in \cref{fig:HadMeasCorrectScheme} will be labeled as $\ket{\psi^{(i)}}$. For instance, the state at the output of $\ket{\overline{H}}_{\text{nf}}$ will be $\ket{\psi^{(1)}}$.

\underline{Case 1: fault in $\ket{\overline{H}}_{\text{nf}}$}.

The output of $\ket{\overline{H}}_{\text{nf}}$ is given by \cref{eq:OutHadNonFT1}. To simplify the notation, let $\overline{E} = \bar{E}^{(x)}_{i}\bar{E}^{(z)}_{j}$, $\widetilde{\overline{E}} = (\overline{H}) \overline{E} (\overline{H})$, $E_p = E^{(x)}_{i}E^{(z)}_{j}$, and $\widetilde{E_p} = \overline{H}E_{p}\overline{H}$. With this notation, the output to the first $H_{m}^{f_2}$ circuit (where we only write the first ancilla since the flag qubits have no effect) can be written as
\begin{align}
\ket{\psi^{(2)}} = \frac{E_p\overline{E}+\widetilde{E_p}\widetilde{\overline{E}}}{2}\ket{\overline{H}}\ket{+} + \frac{E_p\overline{E}-\widetilde{E_p}\widetilde{\overline{E}}}{2}\ket{\overline{H}}\ket{-}
\end{align}

There are several cases to consider
\begin{enumerate}
\item $E_p = I$.

If $\overline{E} \in \{ \overline{X}, \overline{Z} \}$, 

\begin{align}
\ket{\psi^{(2)}} = \frac{1}{2}\ket{\overline{H}}\ket{+} \pm \frac{1}{2}i\ket{-\overline{H}}\ket{-}.
\end{align}
Thus a $\pm1$ measurement outcome gives $\ket{\pm \overline{H}}$ and all three $H_{m}^{f_2}$ circuits will result in a $\pm1$ outcome. If $\overline{E} = \overline{Y}$, $\ket{\psi^{(2)}} = \ket{-\overline{H}}\ket{-}$ and all three $H_{m}^{f_2}$ measurement outcomes will be $-1$.

For the case where all three measurement outcomes are $-1$, applying $\overline{Y}$ to $\ket{\psi_{\text{final}}}$, we will have $\ket{\psi_{\text{final}}} = \ket{\overline{H}}$.

\item $E_p = Y_{i}$.

In this case 
\begin{align}
\ket{\psi^{(2)}} = \frac{1}{2}Y_{i}(\overline{E} - \widetilde{\overline{E}})\ket{\overline{H}}\ket{+} + \frac{1}{2}Y_{i}(\overline{E} + \widetilde{\overline{E}})\ket{\overline{H}}\ket{-}. 
\end{align}
If $\overline{E} \in \{ \overline{X}, \overline{Z} \}$, $\ket{\psi^{(2)}} = \pm \frac{i}{\sqrt{2}}  Y_{i}\ket{-\overline{H}}\ket{+} + \frac{1}{\sqrt{2}}Y_{i}\ket{\overline{H}}\ket{-}$. The $Y_{i}$ error will be corrected by the following EC round. For a $\pm1$ measurement outcome, $\ket{\psi^{(2)}} = Y_{i}\ket{\mp\overline{H}}$ and the next two $H_{m}^{f_2}$ measurements will be $\mp1$ with  $\ket{\psi_{\text{final}}} = \ket{\mp\overline{H}}$. Thus if the three  $H_{m}^{f_2}$ measurement outcomes are $+--$, applying $\overline{Y}$ to $\ket{\psi_{\text{final}}}$ will give $\ket{\psi_{\text{final}}} = \ket{\overline{H}}$. If the measurement outcome is $-++$, $\ket{\psi_{\text{final}}} = \ket{\overline{H}}$. 

If $\overline{E} = \overline{Y}$, $\ket{\psi^{(2)}} =Y_{i}\ket{-\overline{H}}\ket{+}$ and $Y_{i}$ will be corrected by the next EC. Hence the three $H_{m}^{f_2}$ measurement outcomes will be $+--$. Applying $\overline{Y}$ to $\ket{\psi_{\text{final}}}$ will result in $\ket{\psi_{\text{final}}} = \ket{\overline{H}}$.

Lastly, if $\overline{E} = I$, $\ket{\psi^{(2)}} =Y_{i}\ket{\overline{H}}\ket{-}$. The $Y_{i}$ error will be corrected by the next EC and the three $H_{m}^{f_2}$ measurements will be $-++$. 

\item $E_{p} \in \{ X_{i}, Z_{j} \}$.

If $\overline{E} = \overline{Y}$ then 
\begin{align}
\ket{\psi^{(2)}} &= \pm \frac{1}{2}(X_{i} - Z_{i})\ket{-\overline{H}}\ket{+}  \nonumber \\
&+ \frac{1}{2}(X_{i} + Z_{i})\ket{-\overline{H}}\ket{-}.
\end{align}
The three $H_{m}^{f_2}$ measurements will be $\pm --$ with $\ket{\psi_{\text{final}}} = \ket{-\overline{H}}$. Applying $\overline{Y}$ will give the correct state. 

If $\overline{E} \in \{ \overline{X}, \overline{Z} \}$, we have 
\begin{align}
\ket{\psi^{(2)}} &= \frac{1}{2}(X_{i}\overline{Z} + Z_{i}\overline{X}) \ket{\overline{H}}\ket{+} \nonumber \\
&\pm \frac{1}{2}(Z_{i}\overline{X} - X_{i}\overline{Z})\ket{\overline{H}}\ket{-},
\label{eq:Take1}
\end{align}
or 
\begin{align}
\ket{\psi^{(2)}} &= \frac{1}{2}(X_{i}\overline{X} + Z_{i}\overline{Z}) \ket{\overline{H}}\ket{+} \nonumber \\
&\pm \frac{1}{2}(X_{i}\overline{X} - Z_{i}\overline{Z})\ket{\overline{H}}\ket{-}.
\label{eq:Take2}
\end{align}
Since in both \cref{eq:Take1,eq:Take2} the states are linear combinations of products of single-qubit Pauli's and logical operators, after the first EC circuit the output state will collapse to $\ket{\psi^{(3)}} = \overline{E} \ket{\overline{H}}$ where $\overline{E} \in \{ \overline{X}, \overline {Z} \}$ (regardless of whether the first $H_{m}^{f_2}$ measurement outcome was $\pm1$). During the second $H_{m}^{f_2}$ measurement, the output state will become 
\begin{align}
\ket{\psi^{(4)}} &= \frac{\overline{E} + \widetilde{\overline{E}}}{2} \ket{\overline{H}}\ket{+} + \frac{\overline{E} - \widetilde{\overline{E}}}{2} \ket{\overline{H}}\ket{-} \nonumber \\
&= \ket{\overline{H}}\ket{+} \pm i\ket{-\overline{H}}\ket{-}
\label{eq:Take3}
\end{align}

From \cref{eq:Take3} we conclude that the three $H_{m}^{f_2}$ measurement outcomes will either be $\pm ++$ with $\ket{\psi_{\text{final}}} = \ket{\overline{H}}$ or $\pm --$ with $\ket{\psi_{\text{final}}} = \ket{-\overline{H}}$ (in which case we apply $\overline{Y}$).

\item $E_{p} = X_{i}Z_{j}$ with $i \neq j$.

$E_{p}$ will be corrected in the next EC round. A similar calculation as the examples above show that the possible measurement outcomes of the three $H_{m}^{f_2}$ are $+++$, $-++$ and $+--$. If the last two measurements are $-1$, then apply $\overline{Y}$ to the data. In all cases (and after applying the necessary logical operations) $\ket{\psi_{\text{final}}} = \ket{\overline{H}}$. 
\end{enumerate}

\underline{Case 2: fault in the first $H_{m}^{f_2}$ circuit}.

If the fault results in a data-qubit error of weight greater than one, as was shown in \cref{subsec:ErrorDistinguishability}, the possible errors can be distinguished when considering the full six-bit syndrome as well as applying $H_{1}H_{3}H_{4}$ or $I$ to the data depending on the flag outcomes. Thus the output state of the first $H_{m}^{f_2}$ measurement can be expressed as $\ket{\psi^{(2)}} = E_{f}\ket{\overline{H}}\ket{\pm}$ where $E_{f}$ will be corrected by the following EC. Hence the possible measurement outcomes of the three $H_{m}^{f_2}$ circuits are $\pm++$ and the final output state will be $\ket{\psi}_{\text{final}} = \ket{\overline{H}}$.

\underline{Case 3: fault in the first EC}.

The input state to the EC will be $\ket{\overline{H}}$. If a fault results in a data-qubit error of weight greater than one, there will be a flag and the error will be corrected. The output of the EC will be $\ket{\overline{H}}$ and all three $H_{m}^{f_2}$ measurements will be $+1$ with $\ket{\psi}_{\text{final}} = \ket{\overline{H}}$.

If there are no flags, the output state of the EC can be written as  $\ket{\psi^{(3)}} = E_{p}\ket{\overline{H}}$ where $E_{p} = E^{(x)}_{i}E^{(z)}_{j}$. The output state of the second $H_{m}^{f_2}$ measurement will be given by
\begin{align}
\ket{\psi^{(4)}} = \frac{E_{p} + \widetilde{E}_{p}}{2}\ket{\overline{H}}\ket{+} + \frac{E_{p} - \widetilde{E}_{p}}{2}\ket{\overline{H}}\ket{-}.
\end{align}

Note that the error will be corrected in the next EC round. However the type of error can affect the $H_{m}^{f_2}$ measurement outcomes.
\begin{enumerate}
\item $E_{p} = Y_{i}$

In this case $\ket{\psi^{(4)}} = Y_{i}\ket{\overline{H}}\ket{-}$ and the second $H_{m}^{f_2}$ measurement will be $-1$ (with all three measurement outcomes being $+-+$) and $\ket{\psi}_{\text{final}} = \ket{\overline{H}}$\footnote{This case shows why it is important to measure the logical Hadamard operator three times since if the first two measurement outcomes are $+-$, then the data qubit state can either be $\ket{\overline{H}}$ as in this case or $\ket{-\overline{H}}$ as shown in Case 1 (see the text following \cref{eq:Take3}).}.

\item $E_{p} \in \{X_{i}, Z_{i} \}$.

In this case $\ket{\psi^{(4)}} = \frac{X_{i} + Z_{i}}{2}\ket{\overline{H}}\ket{+} \pm \frac{X_{i} - Z_{i}}{2}\ket{\overline{H}}\ket{-}$. Thus the second $H_{m}^{f_2}$ measurement will be $\pm 1$ and all three $H_{m}^{f_2}$ measurements will be $+ \pm +$ with $\ket{\psi}_{\text{final}} = \ket{\overline{H}}$.

\begin{table}
\begin{centering}
\begin{tabular}{|c|c|}
\hline 
$H_{m}^{f_2}$ measurement outcomes & Logical gate correction\tabularnewline
\hline
\hline 
$+++$ & $I$ \tabularnewline
\hline 
$---$ & $\overline{Y}$ \tabularnewline
\hline 
$+--$ & $\overline{Y}$ \tabularnewline
\hline
$-++$ & $I$ \tabularnewline
\hline 
$+-+$ & $I$ \tabularnewline
\hline 
$++-$ & $I$ \tabularnewline
\hline 
\end{tabular}
\par\end{centering}
\caption{\label{tab:LogicAppMeasOut} Logical correction to apply at the end of the circuit in \cref{fig:HadMeasCorrectScheme} given the three measurement outcomes of the $H_{m}^{f_2}$ circuit. The comparison between rows three and five of this table shows why three logical Hadamard measurements are required instead of two.}
\end{table}

\item $E_{p} = X_{i}Z_{j}$ with $i \neq j$.

The analysis of this case is analogous to the case where $E_{p} = Y_{i}$.
\end{enumerate}

\underline{Case 4: fault in the second $H_{m}^{f_2}$ circuit}.

The analysis is identical to Case 2. The possible measurement outcomes of the $H_{m}^{f_2}$ circuits are $+ \pm +$ with $\ket{\psi}_{\text{final}} = \ket{\overline{H}}$.

\underline{Case 5: fault in the second EC circuit}.

The analysis of this case is analogous to Case 3. The possibilities for the three $H_{m}^{f_2}$ measurement outcomes are $++ \pm$ with $\ket{\psi}_{\text{final}} = E_{p}\ket{\overline{H}}$ where $E_{p} = E^{(x)}_{i}E^{(z)}_{j}$. Note however that $E_{p}$ is a correctable error so both criteria in \cref{Def:FaultTolerantPrep} are still satisfied. 

\underline{Case 6: fault in the third $H_{m}^{f_2}$ circuit}.

The analysis is analogous to Case 4. However if there is a flag, an extra EC round should be performed following the third $H_{m}^{f_2}$ circuit. The three $H_{m}^{f_2}$ measurement outcomes will be $++ \pm$ with $\ket{\psi}_{\text{final}} = E_{p}\ket{\overline{H}}$ where $E_{p} = E^{(x)}_{i}E^{(z)}_{j}$ as in Case 5. 

A summary of the logical operations to apply to  $\ket{\psi}_{\text{final}}$ based on the three $H_{m}^{f_2}$ measurement outcomes is given in \cref{tab:LogicAppMeasOut}.

\section{Teleporting into code blocks}
\label{app:GateTeleportation}

 \begin{figure}
	\centering
	\includegraphics[width=0.4\textwidth]{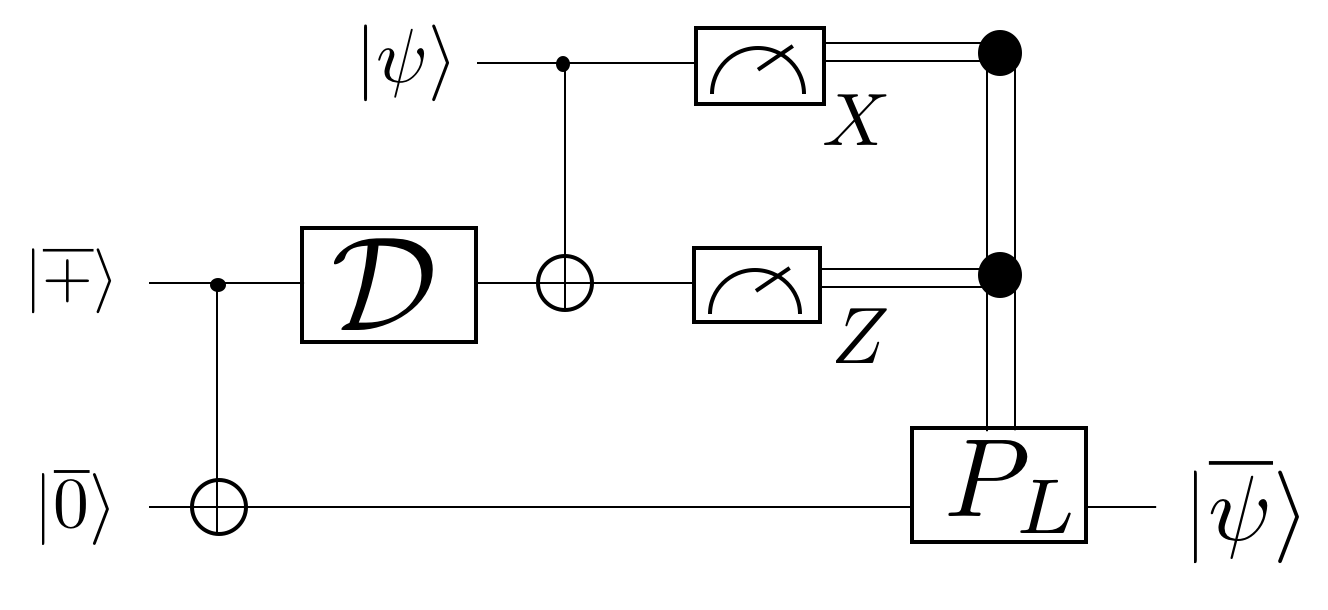}
	\caption{Circuit for teleporting the single-qubit state $\ket{\psi}$ into the codeblock. The operation $\mathcal{D}$ is used to decode one of the code blocks, and $P_{L}$ is a logical Pauli which is applied to complete the teleportation protocol.}
	\label{fig:GateTeleportation}
\end{figure}

In this section we review a general method introduced by Knill for preparing an encoded state $\ket{\overline{\psi}}$ from a physical qubit state $\ket{\psi}$ using teleportation \cite{Knill05}. 

The circuit used to implement the teleportation protocol is illustrated in \cref{fig:GateTeleportation}. After preparing a logical Bell state, one of the code blocks is decoded. A CNOT is applied between $\ket{\psi}$ and the decoded state. After measuring both qubits in the $X$ and $Z$ basis, a logical Pauli operator is applied to the code block in order to complete the teleportation protocol. 

In \cite{Aliferis07}, Aliferis showed that the probability of a logical fault occurring during the teleportation protocol can be bounded by  
\begin{align}
p_{L} \le 3p^{(k)} + p_{\text{dec}}^{(k)} + 4p.
\label{eq:Bound1}
\end{align}
Here we are assuming that the code block is encoded with $k$-levels of concatenation. Assuming a stochastic noise model where encoded gates fail with probability at most $p^{(k)}$,  a fault in the encoded Bell state is upper bounded by $3p^{(k)}$. Since there are four locations in the physical part of the teleportation circuit, and with $p^{(0)} = p$, the probability of this part is bounded by $4p$. Lastly, $p_{\text{dec}}^{(k)}$ is a bound on the probability of failure of the decoding circuit $\mathcal{D}$. 

The level-$k$ block is decoded recursively as follows. The level-$k$ circuit comprises level-$(k-1)$ gates which, when applied to the code block, results in a level-$(k-1)$ encoded state. Then $\mathcal{D}$ is applied again using level-$(k-2)$ gates and so on. Assuming that $\mathcal{D}$ has $D$ locations, we can bound $p_{\text{dec}}^{(k)}$ by
\begin{align}
p_{\text{dec}}^{(k)} \le D\sum_{j=0}^{k-1}p^{(j)},
\label{eq:Bound2}
\end{align}
where $p^{(j)}$ is a bound on the failure probability of level-$j$ gates. 

In this paper, instead of using the bounds in \cref{eq:Bound1,eq:Bound2}, we perform a direct simulation of the teleportation circuit using the methods in \cref{sec:NoiseModelNotation} in order to obtain smaller constant pre-factors (since not all fault locations will lead to a logical fault).

\section{Overhead analysis of the $\ket{H}$ state preparation scheme}
\label{app:HprepOverheadAnalysis}

In this section we provide a detailed description of the qubit and gate overhead analysis for preparing an encoded $\ket{H}$ state using our error detection scheme. 

\subsection{Qubit overhead analysis}
\label{subsec:QubitOverhead}

We begin with a few definitions. Let $p_{A}^{H(l)}$ be the probability that an encoded $\ket{H}$ state at level-$l$ passes the verification test of the circuit in \cref{fig:HadMeasDetectScheme} and let $n_{T}^{H(l)}$ be the total number of qubits used to prepare an encoded $\ket{H}$ state at level-$l$. At the first concatenation level, the largest component of the circuit in \cref{fig:HadMeasDetectScheme} is the EC, which requires 10 qubits in its implementation. Hence we have
\begin{align}
\langle n_{T}^{H(1)}(p) \rangle = \frac{10}{p_{A}^{H(1)}(p)},
\end{align}
for a physical error rate $p$ of the depolarizing noise model. 
 \begin{figure}
	\centering
	\includegraphics[width=0.45\textwidth]{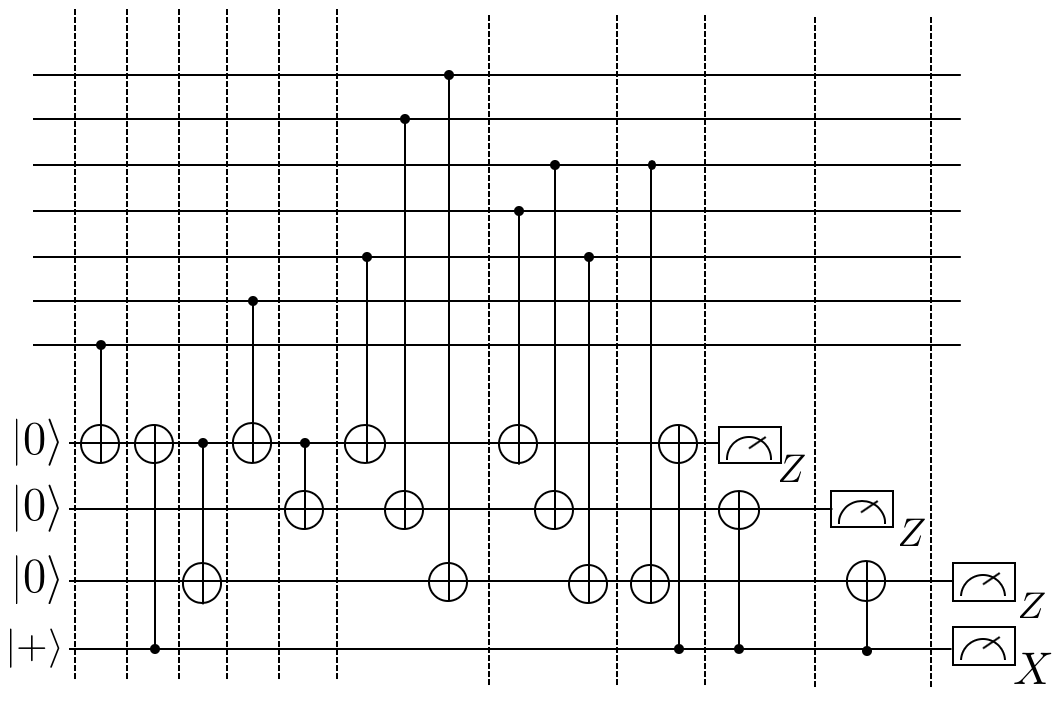}
	\caption{Flag fault-tolerant circuit for measuring the $Z$ stabilizers of the \codepar{7,1,3} Steane code obtained from \cite{CR17v1}. The dashed vertical lines are used to separate different time steps.}
	\label{fig:ZstabMeas}
\end{figure}

At higher levels, a few considerations are necessary when considering different contributions to the qubit overhead. First, for the level-$l$ circuit in \cref{fig:NonFaultTolerantHadCircuit}, it is important to take into account the fault-tolerant preparation of the level-$(l-1)$ $\ket{0}$ and $\ket{+}$ states. The level-$(l-1)$ $\ket{+}$ state is obtained by first preparing the state $\ket{+^{(l-2)}}^{\otimes 7}$ (which is a $+1$ eigenstate of all the $X$-stabilizers), followed by measuring the $Z$-stabilizers using the circuit of \cref{fig:ZstabMeas} (which was shown to be fault-tolerant in \cite{CR17v1}). Depending on the measurement and flag outcomes, it might be necessary to repeat the measurement of $X$ or $Z$ stabilizers without the flag qubit. Note that at the first level, the circuit in \cref{fig:ZstabMeas} requires 11 qubits instead of 10 since there needs to be at least one ancilla qubit prepared in the $\ket{+}$ state in order to detect errors of weight greater than two arising from a single fault. A similar protocol is used to fault-tolerantly prepare a level-$(l-1)$ $\ket{0}$ state.

 \begin{figure}
	\centering
	\includegraphics[width=0.45\textwidth]{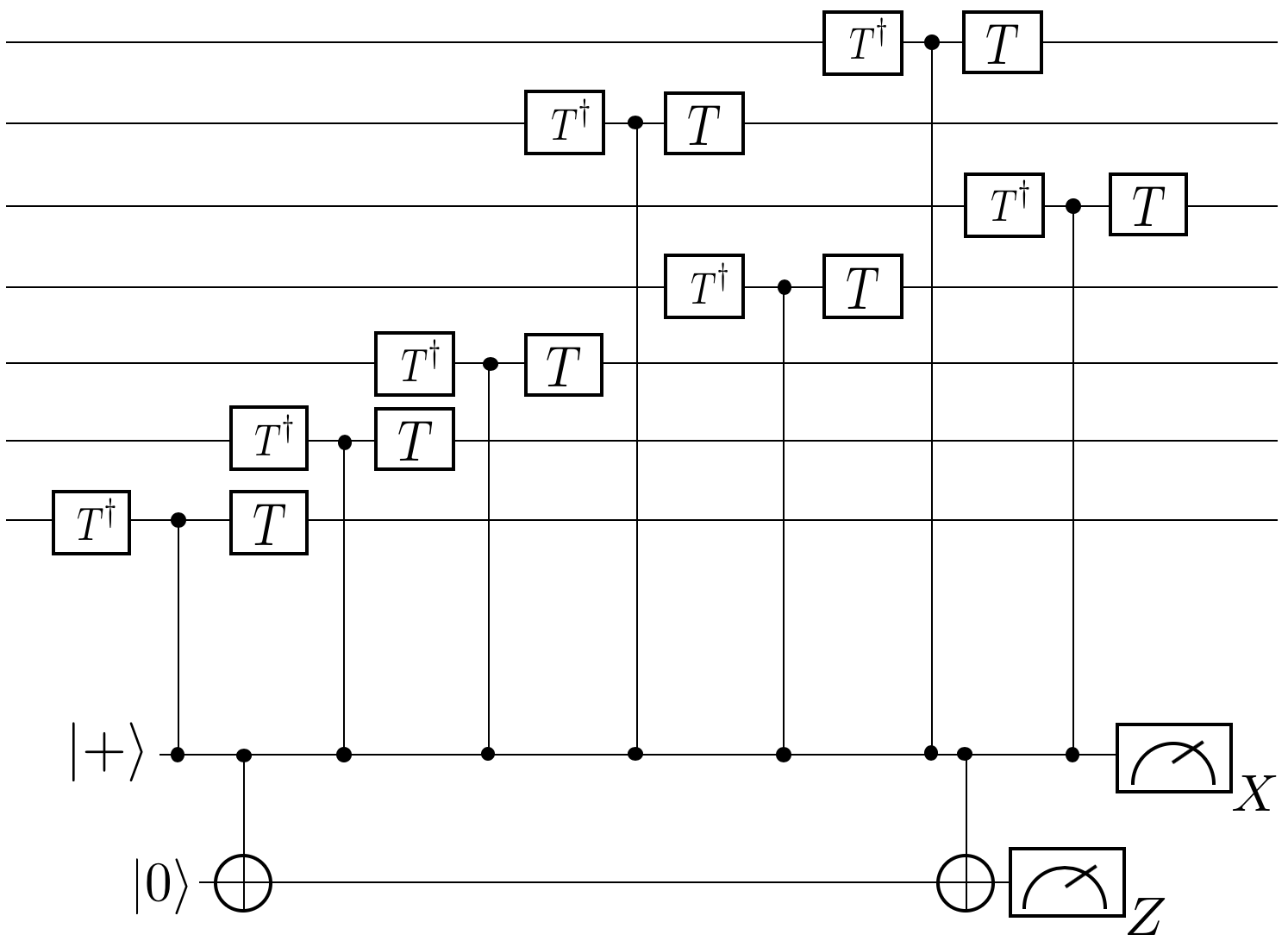}
	\caption{Circuit used to measure the logical Hadamard operator for concatenation levels $l \ge 2$. The $T$ gates are implemented using the circuit in \cref{fig:TgateCirc}. At level-$l$, two level-$(l-1)$ $\ket{H}$ resource states are required for the entire circuit. We assume that the two resource states can be reused for each parallel implementation of $T$ and $T^{\dagger}$.}
	\label{fig:HmeasWithT}
\end{figure}

Next, the details of the implementation of the controlled-Hadamard gates are considered as follows. Since the controlled-Hadamard gates are decomposed as shown in \cref{fig:CHdecomp} with the level-$(l-1)$ $T$ and $T^{\dagger}$ gates implemented using level-$(l-1)$ $\ket{H}$ states as shown in \cref{fig:TgateCirc}, at level $l \ge 2$, the logical Hadamard gate is measured using the circuit in \cref{fig:HmeasWithT}. Due to the way in which we parallelize the the circuit in \cref{fig:HmeasWithT}, only two level-$(l-1)$ resource states are required at each time step, apart from the first and last time step where we only need one resource state. In addition, a level-$(l-1)$ resource state is required for the circuit in \cref{fig:NonFaultTolerantHadCircuit}.

In order to minimize the qubit overhead, we consider preparing in parallel $m^{(1)}_{l} \ge 1$ level-$(l-1)$ resource states at a time step where one resource state is required, and $m^{(2)}_{l} \ge 2$ level-$(l-1)$ resource states at a time step where two resource states are required. At a time step where one level-$(l-1)$ resource state is required, if none of the $m^{(1)}_{l}$ resource states pass the verification test, the protocol is aborted and begins anew. Similarly, at a time step where two resource states are required, we need at least two of the $m^{(2)}_{l}$ resource states to pass the verification test, otherwise, the protocol is aborted. 

Let $p_{AP1}^{(l)}$ be the probability that at least one of the $m^{(l)}_{1}$ level-$(l-1)$ $\ket{H}$ states passes the verification test and $p_{AP2}^{(l)}$ be the probability that at least two of the $m^{(l)}_{2}$ level-$(l-1)$ $\ket{H}$ states pass the verification test. We have that

\begin{align}
p_{AP1}^{(l)}(p) = \sum_{k=1}^{m^{(l)}_{1}} {m^{(l)}_{1} \choose k} (p_{A}^{H(l-1)}(p))^{k} (1-p_{A}^{H(l-1)}(p))^{m^{(l)}_{1} - k},
\label{eq:mlPAP1}
\end{align}
and
\begin{align}
p_{AP2}^{(l)}(p) = \sum_{k=2}^{m^{(l)}_{2}} {m^{(l)}_{2} \choose k} (p_{A}^{H(l-1)}(p))^{k} (1-p_{A}^{H(l-1)}(p))^{m^{(l)}_{2} - k}.
\label{eq:mlPAP2}
\end{align}

\begin{table}
\begin{centering}
\begin{tabular}{|c|c|}
\hline 
 Acceptance probabilities for preparing $m_{1}^{(l)}$ and $m_{2}^{(l)}$ \\
 resource states \tabularnewline
\hline
\hline 
$p_{AP1}^{(2)}(p) \approx 0.999887 - 73.8 p$  \tabularnewline
\hline 
$p_{AP1}^{(2)}(p) \approx 0.99998 - 149 p$  \tabularnewline
\hline 
$p_{AP1}^{(2)}(p) \approx 0.999897 - 86.7 p - 3.71\times 10^{5} p^2$  \tabularnewline
\hline
$p_{AP1}^{(2)}(p) \approx 0.99994 - 177 p - 6.97\times 10^{5} p^2$ \tabularnewline
\hline 
\end{tabular}
\par\end{centering}
\caption{\label{tab:AcceptProbsm1m2} List of the acceptance probabilities $p_{AP1}^{(l)}(p)$ and $p_{AP2}^{(l)}(p)$ at the second and third concatenation levels. The expressions were obtained by solving \cref{eq:mlPAP1,eq:mlPAP2} numerically for values of $p_{AP1}^{(l)}(p),p_{AP2}^{(l)}(p) \in [1-4\times 10^{-2}, 1-10^{-3}]$ increasing in increments of $10^{-2}$ and choosing the pair of values which minimize the total average qubit overhead $\langle n_{T}^{H(l)}(p) \rangle$. We round the coefficients of $p$ and $p^2$ to three digits.}
\end{table}

At each level, we set the values of $p_{AP1}^{(l)}(p)$ and $p_{AP2}^{(l)}(p)$ from $1-10^{-3}$ to $1-4 \times 10^{-2}$ in increments of $10^-2$ and obtain the corresponding values of $m^{(l)}_{1} $ and $m^{(l)}_{2} $ by solving \cref{eq:mlPAP1,eq:mlPAP2}. The final values of $p_{AP1}^{(l)}(p)$ and $p_{AP2}^{(l)}(p)$ chosen to minimize $\langle n_{T}^{H(l)}(p) \rangle$ (see \cref{eq:FinalQubitOverhead} below and \cref{tab:AcceptProbsm1m2}) . Now, the circuits for preparing level-1 $\ket{0}$ and $\ket{+}$ states require 11 qubits. Additionally, three extra $\ket{0}$ and $\ket{+}$ ancilla states are required, two in the circuit of \cref{fig:HmeasWithT} and three for the EC (we assume that the ancilla qubit's used to measure the Hadamard operator can be reused for the EC). Apart for the resource state used to prepare the circuit in \cref{fig:NonFaultTolerantHadCircuit}, the qubits used in the resource states for implementing $T$ and $T^{\dagger}$ gates can be reused at each time step of the circuit in \cref{fig:HmeasWithT}. Since there are three time steps where a single resource state is required and six times steps where two resource states are required, the average number of qubits required to implement our error detection scheme at level-$l$ is given by 

\begin{align}
\langle n_{T}^{H(l)}(p) \rangle = \frac{(\lceil m^{(l)}_{1} \rceil + 2\lceil m^{(l)}_{2} \rceil) \langle n_{T}^{H(l-1)}(p) \rangle  +9(11^{l-1})}{p_{A}^{H(l)}(p) (p_{AP1}^{(l)}(p))^{3}(p_{AP2}^{(l)}(p))^{6}}.
\label{eq:FinalQubitOverhead}
\end{align}

\subsection{Gate overhead analysis}
\label{subsec:GateOverhead}

Since the circuit in \cref{fig:HadMeasDetectScheme} can be decomposed into three parts, the number of gates required to implement our error detection scheme at level-$l$ can be written as\footnote{Note that in this section, all quantities $n_{G}^{(l)}$ should be written as $\langle n_{G}^{(l)} \rangle$ since we are computing average quantities. However, to avoid cluttering in the notation, we omit the brackets.}

\begin{align}
n_{\ket{H}}^{(l)} = \frac{n_{H_{\text{nf}}}^{(l)} + n_{H_{m}^{f_1}}^{(l)} + n_{\text{ED}}^{(l)}}{p_{A}^{H(l)}(p)},
\label{eq:TotalGateOverhead}
\end{align} 
where $n_{H_{\text{nf}}}^{(l)}$ is the number of gates used for preparing the state $\ket{\overline{H}}_{\text{nf}}$ at level-$l$, $n_{H_{m}^{f_1}}^{(l)}$ is the number of gates required to measure the logical Hadamard operator at level-$l$ and $n_{\text{ED}}^{(l)}$ is the number of gates used in the EC circuit of \cref{fig:ReichardtECcircuit} at level-$l$. We used ED instead of EC since the circuit is used in an error detection scheme. It will also be important to analyze the gate overhead of the circuit in \cref{fig:ReichardtECcircuit} when used in an error correction scheme since all level-$l$ ($l \ge 2$) Clifford gates in our circuits consist of extended rectangles where the EC's are used for error correction instead of error detection. 

Performing a gate count of the level-1 circuits in \cref{fig:ReichardtECcircuit,fig:NonFaultTolerantHadCircuit,fig:HmeasWithT}, we find that

\begin{align}
n_{H_{\text{nf}}}^{(1)} = 3(n_{\ket{+}}^{(0)} + n_{\ket{0}}^{(0)}) + n_{\ket{H}}^{(0)} + 11n_{\text{CNOT}}^{(0)} + 6n_{\text{idle}}^{(0)},
\end{align}
\begin{align}
n_{H_{m}^{f_1}}^{(1)} = &2n_{\text{CNOT}}^{(0)} +7(n_{\text{CZ}}^{(0)} + n_{T}^{(0)} + n_{T^{\dagger}}^{(0)}) \nonumber \\
&   + n_{\ket{+}}^{(0)} + n_{\ket{0}}^{(0)} + 191n_{\text{idle}}^{(0)} + n_{Xm}^{(0)} + n_{Zm}^{(0)},
\end{align}
\begin{align}
 n_{\text{ED}}^{(1)} = n_{\text{EC1}}^{(1)} + \langle n_{\text{ED2}}^{(1)} \rangle n_{\text{EC2}}^{(1)},
 \label{eq:lvl1ED}
\end{align}
and

\begin{align}
 n_{\text{EC}}^{(1)} =  \langle n_{\text{REC1}}^{(1)} \rangle n_{\text{EC1}}^{(1)} + \langle n_{\text{REC2}}^{(1)} \rangle n_{\text{EC2}}^{(1)},
 \label{eq:lvl1EC}
\end{align}
where $n_{Zm}^{(l)} = n_{Xm}^{(l)} = 7^{l}$ are the number of $Z$ and $X$-basis measurement locations.

\begin{figure*}
\begin{center}
\subfloat[\label{fig:aEC1EC2rep}]{%
\includegraphics[width = 0.49\textwidth]{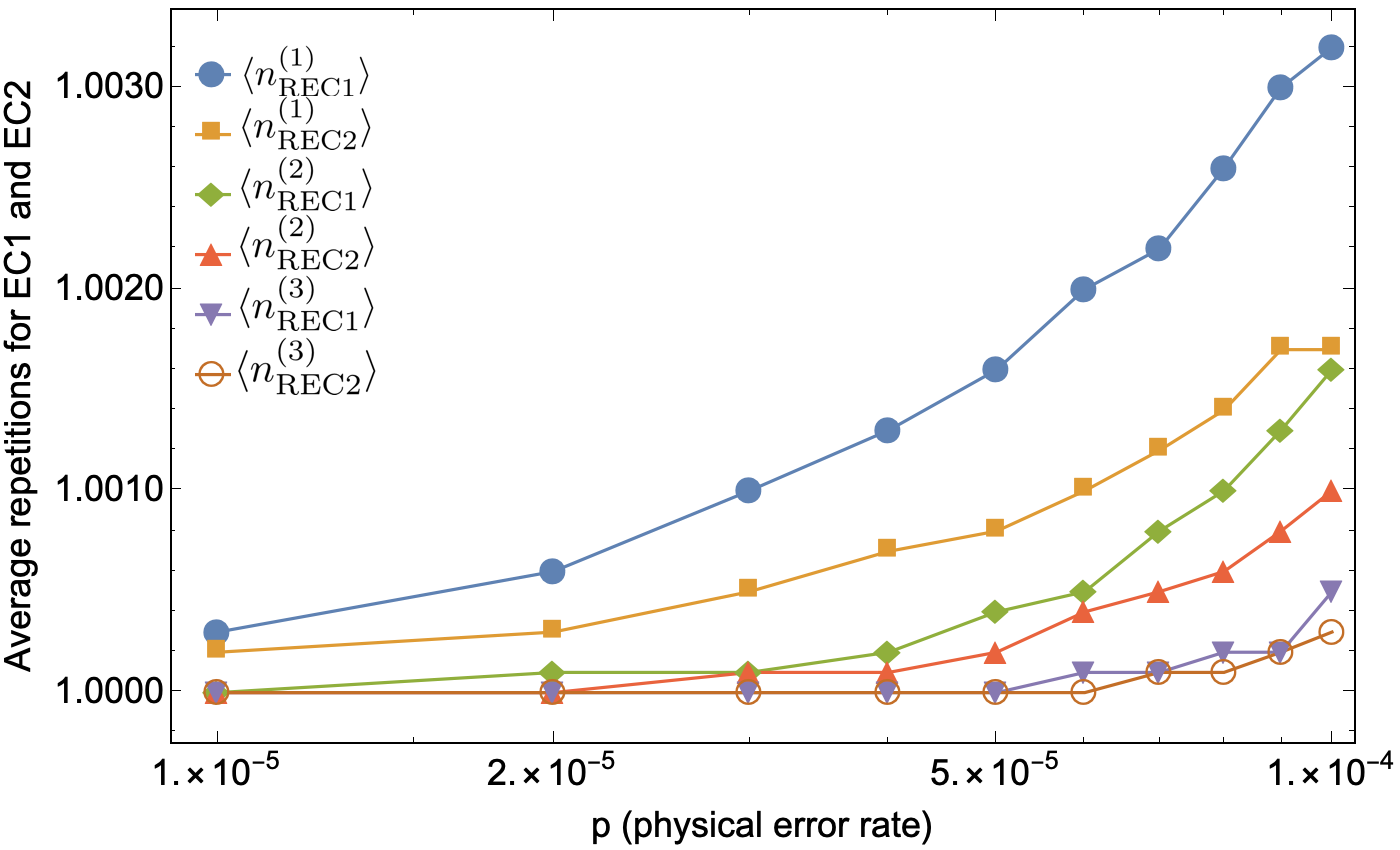}
}\hfill
\subfloat[\label{fig:bOprepRepResults}]{%
\includegraphics[width =0.49\textwidth]{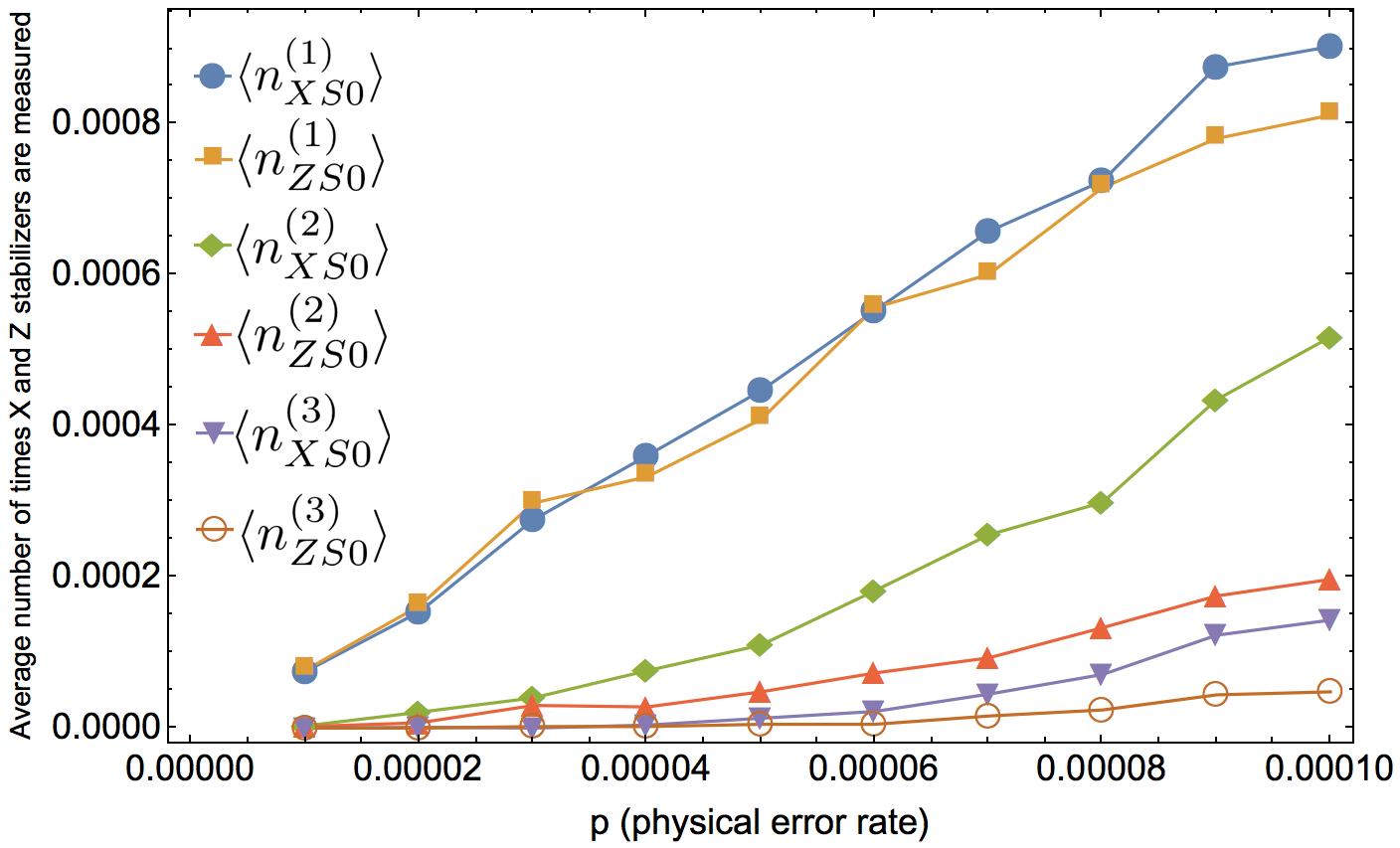}
}\vfill
\subfloat[\label{fig:cPlusprepRepResults}]{%
\includegraphics[width = 0.49\textwidth]{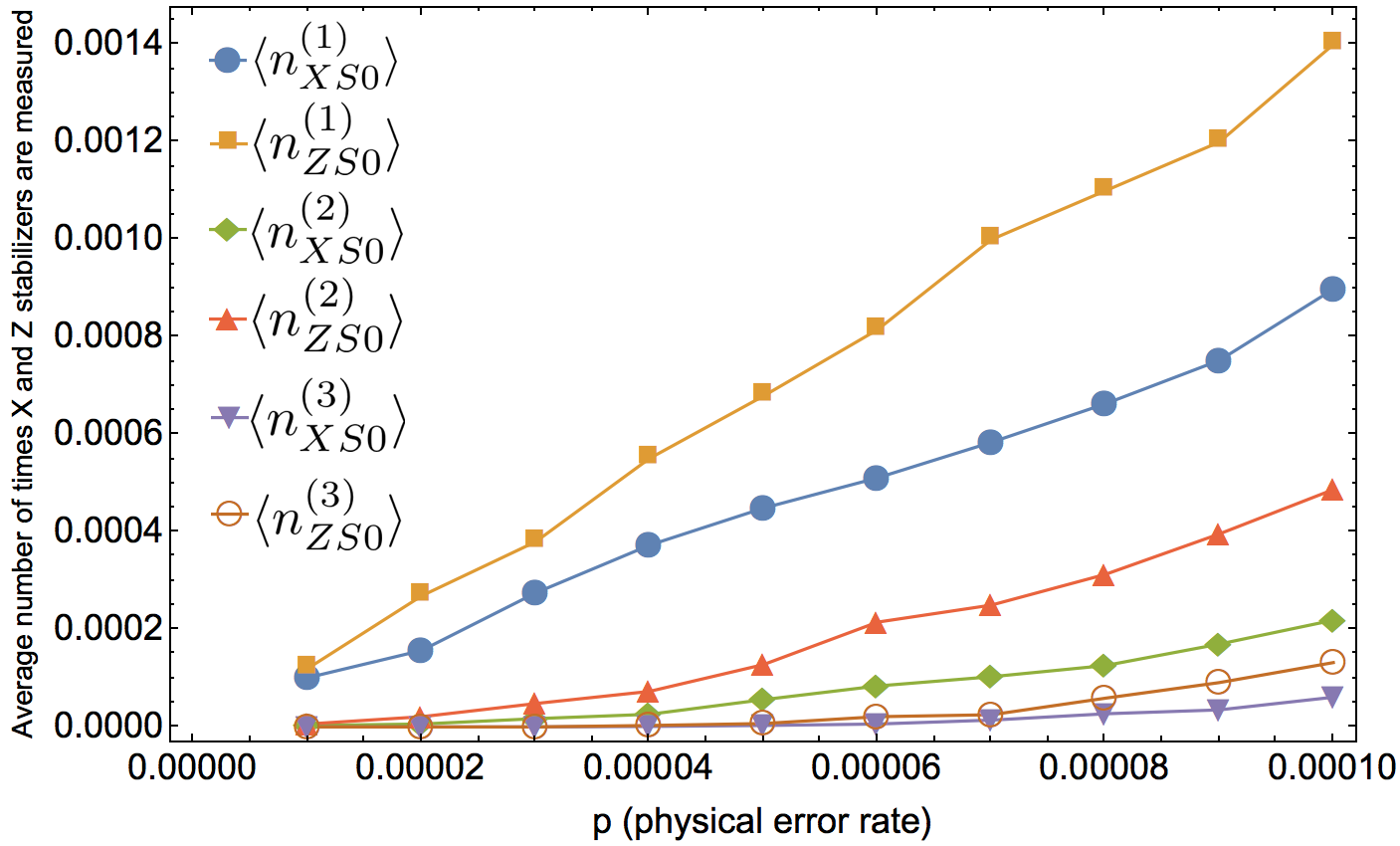}
}\hfill
\subfloat[\label{fig:dEDRepresults}]{%
\includegraphics[width = 0.49\textwidth]{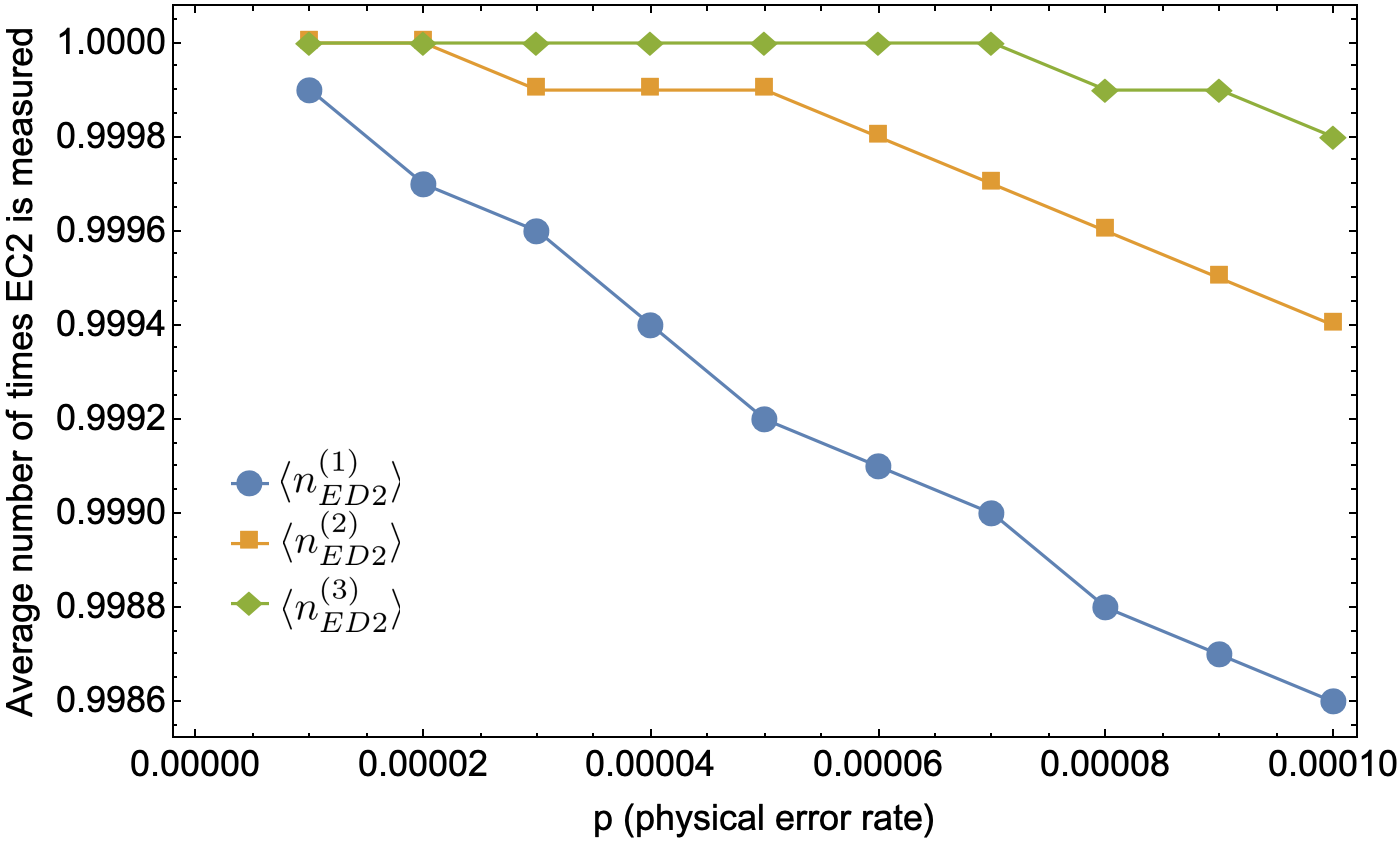}
}
\caption{(a) Plot showing the values obtained for $ \langle n_{\text{REC1}}^{(l)} \rangle$ and $ \langle n_{\text{REC2}}^{(l)} \rangle$ at the first three concatenation levels of the Steane code. (b) Plots showing the values obtained for $\langle n_{XS0}^{(l)} \rangle$ and $\langle n_{ZS0}^{(l)} \rangle$ when preparing the $\ket{0}$ state at the first three concatenation levels. (c) Same as in (b) but for the $\ket{+}$ state. (d) Plot showing the values of $\langle n_{\text{ED2}}^{(l)} \rangle$ at the first three levels. All plots were obtained by performing a Monte-Carlo simulation with $10^{6}$ trials.}
\label{fig:ECRepsAll}
\end{center}
\end{figure*}

In the above, for a gate $G \not\in \{ T, T^{\dagger} \}$, $n_{G}^{(l)} = 7^{l}$ since as we will show below, we will treat the EC's separately from the gates for $l \ge 2$. We split the EC circuit into the two components shown in \cref{fig:ReichardtECcircuit}, which we call EC1 and EC2. As we explained in \cref{subsec:ECcircuit}, depending on the syndrome measurement outcome, the full syndrome measurement can be repeated. Thus $\langle n_{\text{REC1}}^{(l)} \rangle$ and $ \langle n_{\text{REC2}}^{(l)} \rangle$ corresponds to the average number of times that the circuits EC1 and EC2 are implemented at level-$l$. Similarly, $ \langle n_{\text{ED2}}^{(l)} \rangle$ is the average number of times that EC2 is implemented when the EC circuit is used for error detection. These values were obtained through Monte-Carlo simulations with $10^6$ trials (see \cref{fig:ECRepsAll}).

Performing a gate count of the circuits EC1 and EC2, we have that
\begin{align}
n_{\text{EC1}}^{(1)} = 46n_{\text{idle}}^{(0)} + 14n_{\text{CNOT}}^{(0)} + 2(n_{\ket{0}}^{(0)} + n_{Zm}^{(0)}) + n_{\ket{+}}^{(0)} + n_{Xm}^{(0)}
\end{align}
and
\begin{align}
n_{\text{EC2}}^{(1)} = 46n_{\text{idle}}^{(0)} + 14n_{\text{CNOT}}^{(0)} + 2(n_{\ket{+}}^{(0)} + n_{Xm}^{(0)}) + n_{\ket{0}}^{(0)} + n_{Zm}^{(0)}.
\end{align}

Lastly, at concatenation levels $l \ge 2$, the $\ket{0}$ and $\ket{+}$ states are prepared fault-tolerantly using the circuit in \cref{fig:ZstabMeas} to measure the three $Z$ stabilizers of the Steane code, and a similar circuit to measure the three $X$ stabilizers (see the discussion in \cref{subsec:QubitOverhead}). When preparing a logical $\ket{\overline{+}}$ state, if there is a flag in the circuit of \cref{fig:ZstabMeas}, then there could be a $Z$ error of weight greater than one. Thus one must measure the $X$ stabilizers of the Steane code (without using a flag qubit) to correct the $Z$ errors. If there are no flags but the $Z$ stabilizer measurement outcome is nontrivial, then one must repeat the $Z$ syndrome measurement without the flag qubit. A similar analysis holds for preparing a logical $\ket{\overline{0}}$ state. Hence, as in \cref{eq:lvl1ED,eq:lvl1EC}, we must also take into account the average number of times the non-flagged $X$ and $Z$ stabilizers are measured when counting the number of gates required to prepare level-$l$ $\ket{0}$ and $\ket{+}$ states. The averages are obtained by performing a Monte-Carlo simulation (see \cref{fig:ECRepsAll}). In what follows we define $n_{XS}^{(l)}$ and $n_{ZS}^{(l)}$ to be the number of locations in the level-$l$ $X$ and $Z$ stabilizer measurement circuit without the flag qubits and $\langle n_{XS+/0}^{(l)} \rangle $ and $\langle n_{ZS+/0}^{(l)} \rangle $ to be the average number of times these circuits are used when preparing a level-$l$ $\ket{+/0}$ state. Performing the level-1 gate count, we have that

\begin{align}
n_{\ket{+}}^{(1)} = &73n_{\text{idle}}^{(0)} + 15n_{\text{CNOT}}^{(0)} + 8n_{\ket{+}}^{(0)} + 3(n_{\ket{0}}^{(0)} + n_{Zm}^{(0)})  + n_{Xm}^{(0)} \nonumber \\
&  + \langle n_{ZS+}^{(1)} \rangle n_{ZS}^{(1)} + \langle n_{XS+}^{(1)} \rangle n_{XS}^{(1)},
\end{align}
and
\begin{align}
n_{\ket{0}}^{(1)} = &73n_{\text{idle}}^{(0)} + 15n_{\text{CNOT}}^{(0)} + 8n_{\ket{0}}^{(0)} + 3(n_{\ket{+}}^{(0)} + n_{Xm}^{(0)})  + n_{Zm}^{(0)} \nonumber \\
&  + \langle n_{ZS0}^{(1)} \rangle n_{ZS}^{(1)} + \langle n_{XS0}^{(1)} \rangle n_{XS}^{(1)},
\end{align}
with

\begin{align}
n_{ZS}^{(1)} = 48n_{\text{idle}}^{(0)} + 11n_{\text{CNOT}}^{(0)} + 3(n_{\ket{0}}^{(0)} + n_{Zm}^{(0)}),
\end{align}
and
\begin{align}
n_{XS}^{(1)} = 48n_{\text{idle}}^{(0)} + 11n_{\text{CNOT}}^{(0)} + 3(n_{\ket{+}}^{(0)} + n_{Xm}^{(0)}).
\end{align}

 \begin{figure}
	\centering
	\includegraphics[width=0.45\textwidth]{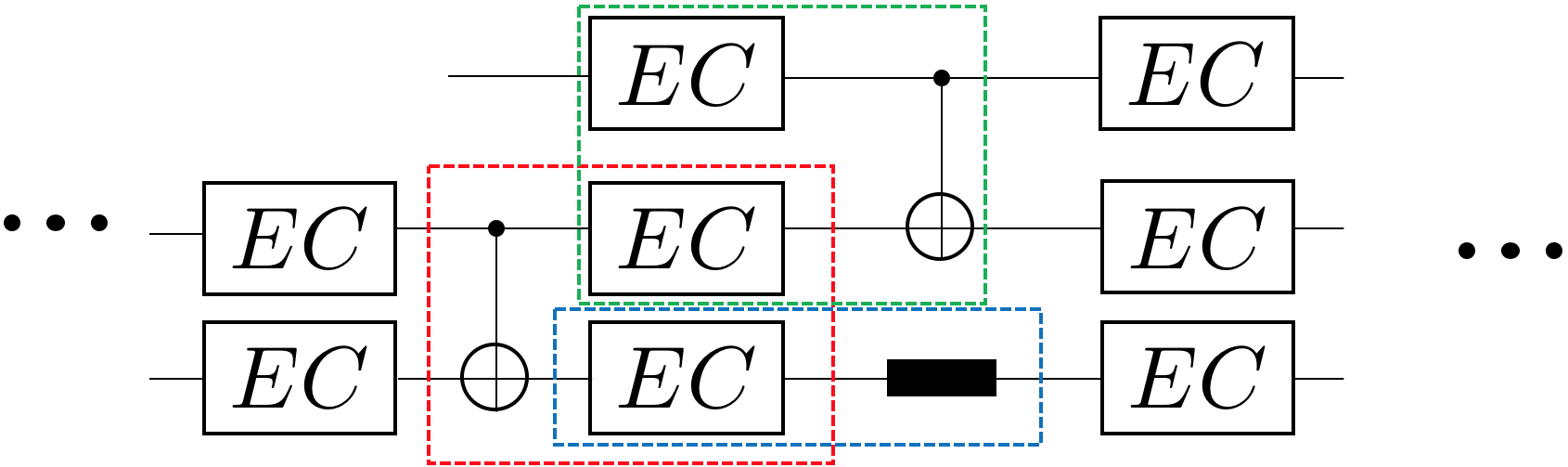}
	\caption{Circuit illustrating shared EC units between two logical CNOT gates and an idle qubit location. It is important not to double-count overlapping EC's when computing the gate overhead at concatenation levels $l \ge 2$.}
	\label{fig:OverlapEC}
\end{figure}

We now have all the tools to obtain the gate overhead at arbitrary concatenation levels. Recall that at concatenation levels $l \ge 2$, all physical gates $G$ in the circuits of \cref{fig:HadMeasDetectScheme} are represented by extended rectangles, which consists of the logical gate $G$ preceded and followed by EC units (in our case, the circuit in \cref{fig:ReichardtECcircuit}). More details on extended rectangles can be found in \cite{AGP06}. When computing the gate overhead of the error detection scheme (\cref{eq:TotalGateOverhead} with $l \ge 2$), we must be careful not to double-count overlapping EC's since consecutive gates will share an EC (see \cref{fig:OverlapEC} for an example). 

Lastly, for $l \ge 2$, the $T$ and $T^{\dagger}$ gates are implemented as shown in \cref{fig:TgateCirc}. We thus see that the overhead for these gates can be computed recursively using

\begin{align}
n_{T}^{(l)} =  n_{T^{\dagger}}^{(l)} = n_{\ket{H}}^{(l-1)}  + 3n_{\text{EC}}^{(l-1)} + 4(7^{l}).
\label{eq:TandTdagFinal}
\end{align}

With the above considerations and using \cref{eq:TandTdagFinal}, the gate overhead can be computed recursively using the following relations. First, the recursive relations for the EC unit are given by
\begin{align}
 n_{\text{EC}}^{(l)} = \langle n_{\text{REC1}}^{(l)} \rangle n_{\text{EC1}}^{(l)} + \langle n_{\text{REC2}}^{(1)} \rangle n_{\text{EC2}}^{(l)},
 \label{eq:lvllEC}
\end{align}
where
\begin{align}
n_{\text{EC1}}^{(l)} = 2n_{\ket{0}}^{(l-1)} + n_{\ket{+}}^{(l-1)} + 73n_{\text{EC}}^{(l-1)} + 63(7^l),
\label{eq:lvlLEC1final}
\end{align}
and
\begin{align}
n_{\text{EC2}}^{(l)} = 2n_{\ket{+}}^{(l-1)} + n_{\ket{0}}^{(l-1)} + 66n_{\text{EC}}^{(l-1)} + 63(7^l).
\label{eq:lvlLEC2final}
\end{align}

Next, the recursive relations for the $\ket{0}$ and $\ket{+}$ states are
\begin{align}
n_{\ket{0}}^{(l)} &=  8n_{\ket{0}}^{(l-1)} +  3n_{\ket{+}}^{(l-1)} + 100n_{\text{EC}}^{(l-1)} + 92(7^l) \nonumber \\
&+ \langle n_{ZS0}^{(l)} \rangle n_{ZS}^{(l)} + \langle n_{XS0}^{(l)} \rangle n_{XS}^{(l)},
\label{eq:OprepfinalLvll}
\end{align}
\begin{align}
n_{\ket{+}}^{(l)} &=  8n_{\ket{+}}^{(l-1)} +  3n_{\ket{0}}^{(l-1)} + 100n_{\text{EC}}^{(l-1)} + 92(7^l) \nonumber \\
&+ \langle n_{ZS+}^{(l)} \rangle n_{ZS}^{(l)} + \langle n_{XS+}^{(l)} \rangle n_{XS}^{(l)},
\label{eq:PlusPrepfinalLvll}
\end{align}
where
\begin{align}
n_{ZS}^{(l)} = 3n_{\ket{0}}^{(l-1)}  + 62(7^l) + 70n_{\text{EC}}^{(l-1)},
\label{eq:NZSFinal}
\end{align}
and
\begin{align}
n_{XS}^{(l)} = 3n_{\ket{+}}^{(l-1)} + 62(7^l) + 70n_{\text{EC}}^{(l-1)}.
\label{eq:NXSFinal}
\end{align}

Using \cref{eq:lvllEC,eq:lvlLEC1final,eq:lvlLEC2final,eq:OprepfinalLvll,eq:PlusPrepfinalLvll,eq:NZSFinal,eq:NXSFinal} and the results in \cref{fig:ECRepsAll}, we can compute the following expressions

\begin{align}
n_{H_{\text{nf}}}^{(l)} = 3(n_{\ket{+}}^{(l-1)} + n_{\ket{0}}^{(l-1)}) + n_{\ket{H}}^{(l-1)} + 28n_{\text{EC}}^{(l-1)} + 17(7^{l-1}),
\end{align}
\begin{align}
n_{H_{m}^{f_1}}^{(l)} = n_{\ket{+}}^{(l-1)} + n_{\ket{0}}^{(l-1)} + 202(7^l) + 14n_{T}^{(l)} + 129n_{EC}^{(l-1)},
\end{align}
and
\begin{align}
 n_{\text{ED}}^{(l)} = n_{\text{EC1}}^{(l)} + \langle n_{\text{ED2}}^{(l)} \rangle n_{\text{EC2}}^{(l)}.
 \label{eq:lvllED}
\end{align}

Note that $n_{\text{EC}}^{0} = 0$ and $n_{T}^{(0)} = n_{T^{\dagger}}^{(0)} = n_{\ket{H}}^{(0)}  = n_{\ket{0}}^{(0)}  = n_{\ket{+}}^{(0)} = 1$.

\section{Overhead analysis for the MEK scheme}
\label{app:MEKOverheadAnalysis}

In this section we provide a detailed description of the qubit and gate overhead analysis required to implement the MEK scheme. 

\subsection{Qubit overhead analysis}
\label{subsec:QubitOverheadMEK}

We first compute the overhead cost of teleporting a physical $\ket{H}$ state to a level-$2$ $\ket{H}$ state and then performing one round of level-$2$ MEK. 

Let $n_{\text{q}ij}^{(T)}$ be the qubit overhead cost of a level-$i$ $\ket{H}$ state teleported to a level-$j$ $\ket{H}$ state. From the teleportation of \cref{fig:GateTeleportation}, a level-$2$ $\ket{0}$ and $\ket{+}$ state must be prepared, each requiring $11^2$ qubits. We assume that these qubits can be reused when performing the logical gates and EC's that follow (since an EC requires only 10 qubits). Including the qubit for the physical $\ket{H}$ state, we have

\begin{align}
n_{\text{q}02}^{(T)} = 2(11^2) + 1 = 243.
\label{eq:Teleport02Qubit}
\end{align}

Next we define $n_{\text{qMEK}}^{(l)}$ to be the qubit overhead cost associated with performing a level-$l$ round of MEK. We also define $a_{\text{MEK}}^{(l)}$ to be the probability that a pair of level-$l$ $\ket{H}$ states are accepted in a level-$l$ round of MEK. When implementing a level-$l$ MEK circuit, we must prepare two level-$l$ $\ket{+}$ states and one level-$l$ $\ket{0}$ state. As was explained in \cref{subsec:QubitOverhead}, we require two additional level-$l$ $\ket{H}$ states in order to perform the $C_{H}$ gates. Thus the level-$2$ MEK circuit qubit overhead with level-$2$ $\ket{H}$ states teleported from physical $\ket{H}$ states is given by

\begin{align}
n_{\text{qMEK}}^{(2)}(p) = \frac{3(11^2) + 4(243)}{a_{\text{MEK}}^{(2)}(p)} = \frac{1335}{a_{\text{MEK}}^{(2)}(p)}.
\label{eq:MEKlvl2qubitCost}
\end{align}

Since the MEK circuit produces two distilled $\ket{H}$ states, when we compare the qubit overhead cost to our $\ket{H}$ state preparation error detection scheme, we will divide \cref{eq:MEKlvl2qubitCost} by two. We note that several additional optimizations are possible. For instance, some of the qubits used for teleporting the $\ket{H}$ states that are used to perform the $C_{H}$ gates could be reused. Additionally, the extra qubits used for preparing the level-$l$ $\ket{0}$ and $\ket{+}$ states for the MEK circuit could also be reused in various parts of the protocol. However these additional optimizations are likely to strongly depend on the particular architecture that is being used and their associated constraints. Therefore, to be as general as possible, these will be omitted. 

We now consider the overhead cost for performing a round of level-$3$ MEK. We first teleport the distilled level-$2$ $\ket{H}$ states to level-$3$ $\ket{H}$ states. We have that

\begin{align}
n_{\text{q}23}^{(T)}(p) &= 2(11^3) + \frac{n_{\text{qMEK}}^{(2)}(p)}{2} \nonumber \\ 
&= 2662 + \frac{1335}{2a_{\text{MEK}}^{(2)}(p)}.
\label{eq:Teleport2to3}
\end{align}

Using the level-$3$ $\ket{H}$ states, the qubit overhead for a level-$3$ MEK simulation is given by

\begin{align}
n_{\text{qMEK}}^{(3)}(p) = \frac{3(11^3) + 4n_{\text{q}23}^{(T)}(p) }{a_{\text{MEK}}^{(3)}(p)},
\label{eq:MEKlvl3qubitCost}
\end{align}
where we will divide \cref{eq:MEKlvl3qubitCost} by two when comparing with the qubit overhead of our error detection scheme.

For the case when a level-$2$ $\ket{H}$ state is prepared using our error detection scheme, we must first teleport the level-$2$ $\ket{H}$ state to a level-$3$ $\ket{H}$ state. Defining $n_{2ED3}^{(T)}$ to be the overhead cost for the teleportation step, we have

\begin{align}
n_{2ED3}^{(T)}(p) = 2(11^3) + \langle n_{T}^{H(2)}(p) \rangle,
\label{eq:2to3TeleportFromCC}
\end{align}
where $\langle n_{T}^{H(2)}(p) \rangle$ is obtained from \cref{eq:FinalQubitOverhead}. Hence the overhead cost for performing a round of level-$3$ MEK is given by

\begin{align}
n_{\text{qMEKED}}^{(3)}(p) = \frac{3(11^3) + 4n_{2ED3}^{(T)}(p)}{2a_{\text{MEKED}}^{(3)}(p)},
\label{eq:MEKlvl3qubitCostED}
\end{align}
where $a_{\text{MEKED}}^{(3)}$ is the acceptance probability for the level-$3$ MEK scheme when the level-$2$ $\ket{H}$ states (teleported to level-$3$) were obtained from our error detection scheme. 

\subsection{Gate overhead analysis}
\label{subsec:GateOverheadMEK}

We now consider the gate overhead analysis for the various rounds of MEK considered in \cref{subsec:QubitOverheadMEK}. 

Let $n_{\text{g}ij}^{(T)}$ be the qubit overhead cost of a level-$i$ $\ket{H}$ state teleported to a level-$j$ $\ket{H}$ state. Two EC's must be performed after applying the logical CNOT gate. We must also take into account the gate cost for the decoding circuit. The decoding circuit requires the preparation of four $\ket{+}$ states and three $\ket{0}$ states. There are eight CNOT gates and the circuit requires a total of 16 EC's. Let $n_{ij}^{(D)}$ be the number of gates in the decoding circuit for level-$i$ to level-$j$ teleportation scheme. We have that

\begin{align}
n_{02}^{(D)} &= 4(n_{\ket{+}}^{(2)} + n_{\ket{+}}^{(1)}) + 3(n_{\ket{0}}^{(2)} + n_{\ket{0}}^{(1)}) \nonumber \\
&+ 16 (n_{\text{EC}}^{(1)} + n_{\text{EC}}^{(2)}) + 8(7^{2} + 7),
\label{eq:GateDecode02}
\end{align}
where $n_{\text{EC}}^{(l)}$, $n_{\ket{0}}^{(l)}$ and $n_{\ket{+}}^{(l)}$ are given by \cref{eq:PlusPrepfinalLvll,eq:OprepfinalLvll,eq:lvllEC}. We then have

\begin{align}
n_{\text{g}ij}^{(T)} = n_{\ket{+}}^{(2)} + n_{\ket{0}}^{(2)} + 2n_{\text{EC}}^{(2)} + n_{02}^{(D)} + 7^{2} + 4.
\label{eq:GateTeleport02}
\end{align}
The last term in \cref{eq:GateTeleport02} is due to the four physical operations of the decoding circuit. 

Next we compute the gate overhead of the level-$2$ MEK scheme. The circuit has 48 logical gates, 177 EC's and requires the preparation of two level-$2$ $\ket{+}$ states and one level-$2$ $\ket{0}$ state. Defining $n_{\text{gMEK}}^{(l)}$ to be the gate overhead of a level-$l$ MEK circuit, we have

\begin{align}
n_{\text{gMEK}}^{(2)} = \frac{4n_{\text{g}02}^{(T)} + 2n_{\ket{+}}^{(2)} + n_{\ket{0}}^{(2)} + 48(7^2)+177n_{\text{EC}}^{(2)}}{a_{\text{MEK}}^{(2)}(p)}.
\label{eq:GateOverheadMEK0to2}
\end{align}
Again when comparing the gate overhead of a level-$2$ MEK circuit to the gate overhead required to prepare a level-$2$ $\ket{H}$ state using our error detection scheme, we will divide \cref{eq:GateOverheadMEK0to2} since an MEK circuit produces two distilled $\ket{H}$ states. 

Following the same steps that lead to \cref{eq:GateOverheadMEK0to2}, it is straightforward to compute the gate overhead for a level-$3$ MEK circuit. Using

\begin{align}
n_{23}^{(D)} &= 4n_{\ket{+}}^{(3)} + 3n_{\ket{0}}^{(3)} + 16 n_{\text{EC}}^{(3)} + 8(7^{3}),
\end{align}
and
\begin{align}
n_{\text{g}23}^{(T)} &=  n_{\ket{+}}^{(3)} + n_{\ket{0}}^{(3)} + 2n_{\text{EC}}^{(3)} \nonumber \\
&+3n_{\text{EC}}^{(2)} + 7^3+7^2 + n_{23}^{(D)} + \frac{n_{\text{gMEK}}^{(2)}}{2},
\label{eq:GateTeleport23}
\end{align}
we get
\begin{align}
n_{\text{gMEK}}^{(3)} = \frac{4n_{\text{g}23}^{(T)} + 2n_{\ket{+}}^{(3)} + n_{\ket{0}}^{(3)} + 48(7^3)+177n_{\text{EC}}^{(3)}}{a_{\text{MEK}}^{(3)}(p)}.
\label{eq:GateOverheadMEK2to3}
\end{align}

The gate overhead for a level-$3$ MEK scheme where a level-$2$ $\ket{H}$ state was prepared from our error detection scheme is obtained as follows. First we compute the gate overhead to teleport the level-$2$ $\ket{H}$ state to a level-$3$ $\ket{H}$ state. It is given by

\begin{align}
n_{\text{g}23\text{ED}}^{(T)}(p) &= n_{\ket{H}}^{(2)} + n_{\ket{+}}^{(3)} + n_{\ket{0}}^{(3)} \nonumber \\
&+ 2n_{\text{EC}}^{(3)} +3n_{\text{EC}}^{(2)} + 7^3+7^2 + n_{23}^{(D)},
\label{eq:GateTeleport23CC}
\end{align}
where $n_{\ket{H}}^{(2)}$ is obtained from \cref{eq:TotalGateOverhead}. Using \cref{eq:GateTeleport23CC} we obtain 

\begin{align}
n_{\text{gMEKED}}^{(3)} = \frac{4n_{\text{g}23\text{ED}}^{(T)}(p) + 2n_{\ket{+}}^{(3)} + n_{\ket{0}}^{(3)} + 48(7^3)+177n_{\text{EC}}^{(3)}}{a_{\text{MEKED}}^{(3)}(p)},
\end{align}
where $a_{\text{MEKED}}^{(3)}$ is the acceptance probability of the level-$3$ MEK circuit when the level-$2$ $\ket{H}$ states (teleported to level-$3$) are obtained from our error detection scheme. 

\section{State vector simulations}
\label{app:StateVectorSim}

Here we describe how we simulate each $|H\rangle$ state preparation and MEK magic state distillation circuit. A common method for simulating fault-tolerant error-correction is to apply the Gottesman-Knill theorem to track how Pauli errors propagate through stabilizer circuits. However, unlike error-correction, circuits for magic state preparation and distillation contain non-Clifford gates that map Pauli errors outside the Pauli group, complicating the analysis. Since all of the circuits we consider act on relatively few qubits, we use the Qiskit state vector simulation \cite{Qiskit} to accurately track error propagation through non-Clifford gates.

Each gate, preparation, idle and measurement location is subject to Pauli channel errors whose probabilities are functions of the physical error rate $p$ that are determined by the noise model at level-1 and the logical error probabilities of fault-tolerant gates at level-2 and above (see \cref{sec:NoiseModelNotation}). For each value of $p$, we run between $10^6$ and $10^8$ Monte-Carlo samples, where each sample draws an error at each fault location. The circuits are composed with an ideal decoder so that the output state is one or two physical qubits for state preparation or distillation, respectively. The output for each sample $i$ is a pure state vector $|\psi_i\rangle$ and a measurement record $m_i$ on which we post-select. 

For simulations of $|H\rangle$ state preparation, we compute overlaps $\langle\psi_i|P|\psi_i\rangle$ for each Pauli $P$ to determine the logical error class, since in this case the logical error is always of this form after ideal decoding. However, for the MEK distillation protocol there are additional failure modes. For example, the two output qubits can be in a maximally entangled state. This can be seen by placing $Y$ errors on the fourth and seventh $|H\rangle$ states, which corresponds to failures of the first $T$ gate and third $T^\dagger$ gate. Therefore, we solve for Pauli channel parameters using an estimate of the output density matrix $\tilde{\rho}\approx \sum_i |\psi_i\rangle\langle\psi_i|$. First, we compute the reduced state $\rho=\mathrm{Tr}_2 \tilde{\rho}$ of one output qubit. Next, we model the output state as a Pauli channel applied to the state $\rho_H=|H\rangle\langle H|$,
\begin{align}
{\cal E}(\rho_H) & = (1-p_x-p_z)\rho_H + p_x X\rho_H X + p_z Z\rho_H Z \\
& = \frac{1}{2\sqrt{2}}\left(\begin{array}{cc}
1+\sqrt{2}-2p_x & 1-2p_z \\
1-2p_z & -1+\sqrt{2}+2p_x\end{array}\right).
\end{align}
For the parameter ranges we considered, we can solve for the Pauli channel parameters in terms of the matrix elements of $\rho$,
\begin{align}
p_x & = \frac{1+\sqrt{2}-2\sqrt{2}\rho_{00}}{2} \\
p_z & = \frac{1-2\sqrt{2}\rho_{01}}{2}.
\end{align}
We substitute these parameters back into the model and verify the density matrices are equal to machine precision.

The logical error probabilities and rejection probabilities are fit to functions of the physical error $p$. State preparation results are summarized in Table~\ref{tab:PrepSimulationResults} and distillation results in Table~\ref{tab:DistillSimulationResults}. For the second round of distillation, the $\mathcal{O}(p^4)$ contribution to the logical error is too small to observe with $10^8$ Monte-Carlo trials, so we do two such simulations. The first case uses ideal $|H\rangle$ states and noisy stabilizer operations, while the second case uses noisy $|H\rangle$ states and ideal stabilizer operations. We simulate the second case at higher physical error rates to estimate the coefficient of the $\mathcal{O}(p^4)$ term. Finally, we approximate the logical error rate and acceptance probability of the noisy circuit as a piecewise function (see the caption of Table~\ref{tab:DistillSimulationResults}).

\begin{table*} 
	\begin{centering}
		\begin{tabular}{|l|c|c|}
			\hline 
			Concatenation & $\textrm{Pr}[\textrm{accept}]$ & $\textrm{Pr}[\textrm{mal}^{(l)}_E|G, p]$ \\
			\hline\hline
			$l=1$ & $(1-p)^{75}$ & $(9.95, 4.41, 7.87) p^2$ \\
			\hline
			$l=2$ & $(1-3000 p^2)^{200}$ & $(1.26, 0.0627, 1.09)\times 10^9 p^4$ \\
			$l=2^\ast$ & $1 - 3.67 p - 2.67\times 10^{3} p^2$ & $(5.14, 0.33, 2.26)\times 10^4 p^4$ \\
			\hline
			$l=3$ & $1 - 84.3 p + 1.60\times 10^{6} p^2- 9.41\times 10^{9} p^3$ & $(1.23, 0.0555, 1.16)\times 10^{24} p^8$ \\
			$l=3^\ast$ & $1 - 84.3 p + 1.60\times 10^{6} p^2 - 9.41\times 10^{9} p^3$ & $(1.24, 0.0683, 1.17)\times 10^{24} p^8$ \\
			\hline
		\end{tabular}
		\par\end{centering}		
	\caption{\label{tab:PrepSimulationResults} State vector simulation results for $|H\rangle$ state preparation circuits using error-detection. The error probabilities $(p_x, p_y, p_z)$ are given as functions of $p$ for each error type $E\in \{X, Y, Z\}$. The entires marked with $(\ast)$  use error-detection circuits for level-1 stabilizer operations (only in the $\ket{H}$ state preparation circuit). The acceptance probabilities are for the level-$l$ circuit and do not include rejections at lower levels of concatenation. The expressions are all approximations computed using a standard non-linear fitting method, and coefficients are rounded to three digits. The level-3 expressions are valid for $p\leq 4\times 10^{-4}$. We notice that both level-3 error polynomials are similar. This is due to the fact that errors arising from the level-2 $\ket{H}$ states have a much smaller contribution to the overall logical error rate compared to the Clifford operations. }
\end{table*}

\begin{table*} 
	\begin{centering}
		\begin{tabular}{|l|c|c|}
			\hline 
			Circuit & $\textrm{Pr}[\textrm{accept}]$ & $\textrm{Pr}[\textrm{mal}^{(l)}_E|G, p]$ \\
			\hline\hline
			hybrid, $l=3$ & $1 - 3.45\times 10^{20} p^6$ & $(0.812, 1.68)\times 10^{26} p^8$ \\
			\hline
			round-1, $l=2$ & $1 - 32.2 p + 2.24\times 10^{4} p^2 - 2.37\times 10^{9} p^3$ & $(150, 152) p^2 + (1.01, 2.00)\times 10^{11}p^4$ \\
			round-2, $l=3$, ideal $|H\rangle$ & $1 - 6.50\times 10^{27} p^8$ & $(1.43, 2.43)\times 10^{26} p^8$ \\
			round-2, $l=3$, ideal stabilizer$^\ast$ & $1 - 1.47\times 10^{3} p^2$ & $(1.97, 1.84)\times 10^{5} p^4$ \\
			round-3, $l=3$ & $1 - 8.00\times 10^{27} p^8$ & $(2.00, 3.04)\times 10^{26} p^8$ \\
			\hline
		\end{tabular}
		\par\end{centering}		
	\caption{\label{tab:DistillSimulationResults} State vector simulation results for Meier-Eastin-Knill distillation circuits. The error probabilities $(p_x, p_z)$ are given as functions of $p$ for each error type $E\in \{X, Z\}$. The expressions are all approximations computed using a standard non-linear fitting method, and coefficients are rounded to three digits. The expressions are valid for $p\leq 2\times 10^{-4}$, except for $(\ast)$ which is valid for $p\leq 5\times 10^{-3}$. For round-2, we combine the two results in the table into the following optimistic piecewise approximation: $\textrm{Pr}[\textrm{accept}]\approx 1 - \mathrm{max}(1.47\times 10^{3} p^2, 6.50\times 10^{27} p^8)$ and $\textrm{Pr}[\textrm{mal}^{(l)}_E|G,p]\approx  (1.97, 1.84)\times 10^{5} p^4 + (1.43, 2.43)\times 10^{26} p^8$.}
\end{table*}

\end{document}